# The Physics of Communicability in Complex Networks

Ernesto Estrada[1]

Department of Mathematics and Statistics, Department of Physics, SUPA, and Institute of Complex Systems, University of Strathclyde, Glasgow G1 1XQ, UK.

Naomichi Hatano

Institute of Industrial Science, University of Tokyo, Komaba, Meguro, Tokyo 153-8505, Japan.

Michele Benzi

Department of Mathematics and Computer Sciences, Emory University, Atlanta, Georgia 30322, USA

---

[1] Corresponding author. E-mail: ernesto.estrada@strath.ac.uk

# ABSTRACT

A fundamental problem in the study of complex networks is to provide quantitative measures of correlation and information flow between different parts of a system. To this end, several notions of communicability have been introduced and applied to a wide variety of real-world networks in recent years. Several such communicability functions are reviewed in this paper. It is emphasized that communication and correlation in networks can take place through many more routes than the shortest paths, a fact that may not have been sufficiently appreciated in previously proposed correlation measures. In contrast to these, the communicability measures reviewed in this paper are defined by taking into account all possible routes between two nodes, assigning smaller weights to longer ones. This point of view naturally leads to the definition of communicability in terms of matrix functions, such as the exponential, resolvent, and hyperbolic functions, in which the matrix argument is either the adjacency matrix or the graph Laplacian associated with the network.

Considerable insight on communicability can be gained by modeling a network as a system of oscillators and deriving physical interpretations, both classical and quantum-mechanical, of various communicability functions. Applications of communicability measures to the analysis of complex systems are illustrated on a variety of biological, physical and social networks. The last part of the paper is devoted to a review of the notion of locality in complex networks and to computational aspects that by exploiting sparsity can greatly reduce the computational efforts for the calculation of communicability functions for large networks.

CONTENTS

# I. INTRODUCTION

## A. Overview: interaction and correlation

In social system studies it is frequently found that agents belonging to the same group tend to behave similarly (Manski, 2000). Social scientists use the term '*interaction*' to explain this empirical regularity and use terms such as "social norms", "peer influences", "neighborhood effects", "conformity", "imitation", "contagion", "epidemics", "bandwagons" or "herd behavior" to refer to them (Merton, 1957; Granovetter, 1979; Manski, 2000). In physics, one is often warned (but not as often heard) that *interaction* and *correlation* are two different concepts. Consider a solid, for example. Each atom in a solid interacts with neighboring atoms mostly, and perhaps with next- and second-next neighboring ones at most. However, if we hit one end of a solid bar, the effect of the action propagates to the other end, which is a manifestation of the fact that an atom on the one end of solid is *correlated* with an atom on the other end. Correlation is indeed *the* driving force of most phase transitions of many-body systems. When a substance undergoes successive phase transitions from gas to liquid on to solid, the interaction range of each atom does not change much but the correlation range grows singularly and becomes macroscopic eventually when the body is solidified. It is not difficult then to realize that the term 'interaction' widely used in social sciences actually refers mostly to the '*social correlation*' that is produced by the networked characteristic of social systems (for a review on the statistical physics of social dynamics see: Castellano *et al.*, 2009). One illustrative example of macroscopic (global) correlation is the cell-phone adoption during the 1990s; it was a "contagion" effect that induced many to buy phones simply because their friends and colleagues were buying them, which then propagated in a correlated way across Europe (Michard and Bouchard, 2005) and the world.

The atomic/social metaphor has been widely used in both social and physical sciences. Some of the pioneers of statistical physics, like Maxwell and Boltzmann, were inspired by the works of social scientists Buckle and Quetelet; see (Ball, 2004). More recently, the metaphor of a 'social atom' and the tools of statistical mechanics were used to explain the structure and dynamics of social and economic systems (Buchanan, 2007), giving rise to the fields of socio- and econophysics (Mantegna and Stanley, 1999; Chakrabarti *et al.*, 2006). This analogy functions very well when the properties studied depend mainly on the networked structural properties of the system analyzed. A good example is a linear chain in which the $i$ th node is connected to the $(i+1)$ th one. It has been proven that a purely one-dimensional system never becomes a solid (Peierls, 1936; Ruelle, 1969). It is roughly explained as follows. In one dimension, correlation could grow only along the chain. A disturbance at only one point of the chain, destroying correlation locally, results in the global destruction of the correlation between the atoms on both ends. However, social networks are much more complex than a linear chain and the analogy with a three-dimensional atomic system is more illuminating. In the three-dimensional system, we know very well that the global correlation can overcome the local disturbance, such as the removal of one atom, to be a solid at low temperatures. This is because there are many more paths along which the correlation can grow than in one dimension. In other words, there is a topologically more complex networked structure in these systems and the complexity of the structure promotes the growth of correlation throughout the systems.

It is clear that not only atomic and social systems display network-like structures. Complex networks are also ubiquitous in many biological, ecological, technological, informational, and infrastructural systems (Albert and Barabási, 2002; Barabási and Oltvai, 2003; Caldarelli, 2007; da Fontoura Costa *et al.*, 2011; Dorogovtsev and Mendes, 2003; Newman, 2003; 2010; Strogatz, 2001; Watts, 2003). A complex network is a representation of a complex system in which the nodes $v_i \in V, i = 1\ldots, n$ of a graph $G = (V, E)$ represent the entities of the systems and the interactions between pairs of entities are accounted for by the links $\{v_i, v_j\} \in E$ of the graph. (In graph theory, a node is also called a vertex and a link an edge.) Consequently, global correlation effects are also observed in a variety of complex networks. For instance, it is well documented that the extinction of one species in an ecological system produces a "cascade" of effects that propagate well beyond the nearest neighbors of the extinguished species (Dunne *et al.*, 2002; Jordán and Scheuring, 2002). In a biological cell where protein-protein interactions form biomolecular networks, it is well documented that 'perturbing' one protein can trigger a cascade of effects that change or modify the synthesis and folding of several other proteins not necessarily directly interacting with the targeted one (Zotenko *et al.*, 2008). For infrastructural network scenarios, the cascade of local failures in power grids that have produced mass blackouts such as the one in eleven USA states and two Canadian provinces on 14[th] August 2003 affecting 50 million people, or those in London, U.K., Sweden-Denmark, and Italy, are palpable examples of correlation effects in complex networks (Makarov *et al.*, 2005). Finally, the propagation of crisis effects in a world-wide networked economy (Count and Bouchard, 2000; Eguíluz and Zimmermann, 2000) alerts us about the importance of the study of correlation in complex networks as a tool of great relevance to understand the structure and functioning of many complex systems in nature and society. It is then essential to understand what topology indeed promotes the growth of correlation and what disturbs it in such systems.

Unfortunately, the term 'correlation' is frequently used in other sciences under different meanings than the one in physics. For instance, the term "correlated effects" is used in social sciences to refer to "interactions" in which "agents in the same group tend to behave similarly because they have similar individual characteristics or face similar institutional environments" (Manski, 2000). In other contexts it mainly refers to the linear interdependence of two or more variables in the statistical sense as measured by a correlation coefficient. Subsequently, in the context of complex networks we have proposed the use of the term "communicability" to refer to the situations in which a perturbation on one node of the network is 'felt' by the rest of the nodes with different intensities. The concept of network communicability was introduced by Estrada and Hatano (2008). The intuition behind this concept is that in many real-world situations the communication between a pair of nodes in a network does not take place only through the optimal shortest-path routes connecting both nodes, but through all possible routes connecting both nodes, the number of which can be enormous in the complex topology of the systems. The information can also go back and forth before arriving at the end node of a given route. The network communicability quantifies such correlation effects in the communication between nodes in complex networks. The most important point that we would like to stress in the present paper is that the number of routes along which the correlation can grow is crucial in the analyses of the structures of complex networks.

There have been other proposals of indices for the communication through complex networks. These indices are mainly used to quantify the self-communicability of a given node in the form of a centrality measure. The most characteristic of these indices are the closeness (Freeman, 1979) and betweenness centrality (Freeman, 1979) and some of their modifications like the information centrality (Stephenson and Zelen, 1989), and betweeness accounting not only for shortest paths (Freeman *et al.*, 1991; Newman 2005; Estrada *et al.*, 2009). In some way the eigenvector centrality introduced by Bonacich (1972; 1989), which is the principal eigenvector of the adjacency matrix of the network, can be considered as a self-communicability function. In this context, we can consider the number $N_k(i)$ of walks of length $k$ starting at node $i$ (see further for proper definition) of a non-bipartite connected network. If $s_k(i) = N_k(i) \cdot \left[ \sum_{j=1}^{n} N_k(j) \right]^{-1}$, for $k \to \infty$, the vector $\left[ s_k(1), s_k(2), \cdots s_k(n) \right]^T$ tends towards the eigenvector centrality (Cvetković *et al.*, 1997). This means that the eigenvector centrality of node $i$ represents the ratio of the number of walks of length $k$ that departs from $i$ to the total number of walks of length $k$ in a non-bipartite connected network when the length of these walks is sufficiently large.

In the present review, we will analyze four other kinds of communicability indices. It is important to note that it is not a matter of deciding which index is the 'correct' one to indicate the communication. There is indeed no standard that we can refer to in judging the 'correctness' of an index. It is a matter of which index is more appropriate to a specific problem than others. In judging the appropriateness, we will have to resort to our intuition and experience. Typically, we would make predictions from various indices and compare them with the result of analyzing actual datasets or sometimes even with a plausible guess. In the present review, we therefore show various specific examples where one index is more appropriate than others.

**B. Correlation function**

Correlation effects can be quantified by the correlation function. The definition of the correlation function depends on the problem under study, but the general idea is to measure how a tiny disturbance at one point of the system propagates to another point of the system; hence the aliases, the propagator and the Green's function. For a general reference on this topic the interested reader is referred to Landau and Lifshitz (1980). In quantum mechanics, the system with the Hamiltonian $H$ evolves in time according to the time-evolution operator

$$U(t) = \exp\left(-\frac{i}{\hbar} H t\right). \tag{1}$$

(For some readers, the form

$$G(E) = (E - H)^{-1}, \quad (2)$$

may be more familiar. Indeed, Eq. (2) is simply the one-sided Fourier transform of Eq. (1) (Sakurai, 1985).) Then, the typical definition of the propagator may be given as

$$C(r,t) = \langle \text{vac} | a_r U(t) a_0^\dagger | \text{vac} \rangle, \quad (3)$$

where $|\text{vac}\rangle$ denotes the vacuum state and $a_0^\dagger$ denotes the particle creation operator at the origin 0, whereas $a_r$ denotes the particle annihilation operator at the distance $r$ from the origin. Equation (3) describes how the impact of creating a particle at the point 0 propagates over the distance $r$ and affects the point $r$ after the time $t$. Obviously, the correlation is strong if there are many paths on which the effect can propagate.

In equilibrium statistical physics, the thermal disturbance is important. Instead of the real-time dynamics in Eq. (3), we then often consider the thermal correlation function, or the thermal Green's function, which may be given in the form

$$C(r,\beta) = \langle \text{vac} | a_r \rho(\beta) a_0^\dagger | \text{vac} \rangle, \quad (4)$$

where

$$\rho(\beta) = Z^{-1} \exp(-\beta H), \quad (5)$$

is the density operator of the Gibbs equilibrium distribution, where $Z = tr[\exp(-\beta H)]$ is the partition function. We will take full advantage of the thermal Green's function (4) throughout the paper. It describes the propagation of disturbance through the system in a thermal bath at the inverse temperature $\beta = 1/kT$, where $k$ here is the Boltzmann constant. The thermal Green's function (4) is indeed the analytic continuation of the Green's function (3) onto the imaginary time axis $it/\hbar = \beta$.

**C. Plan of the article**

We will define the communicability in Sec. II, presenting several definitions. Section III elaborates the analogy between the communicability of complex networks and the correlation of physical systems. We will show that classical and quantum statistical calculations with the adjacency-based and Laplacian-based models result in four different versions of the communicability. Then in Sec. IV we compare the four versions in two specific examples. Section V presents a variety of usages of the communicability in analyses of various complex networks, namely analyses at the microscopic level, the mesoscopic level, the macroscopic level, and the multi-scale level. We also present an interesting application of the communicability with a negative temperature to the analysis of bipartite networks. We discuss the locality of the communicability in Sec. VI, showing instances of exponential decay and slow decay of the communicability, or the correlation function. We finally review recent advances in computing the communicability of large networks. It is a heavy task to compute the communicability as an exponentiated operator. Taking advantage of the sparsity of the adjacency matrix can greatly improve the computational efficiency. The final section is devoted to conclusions.

## II. COMMUNICABILITY IN NETWORKS
### A. Combinatorial definition

In this section we introduce the concept of communicability in networks by using a graph-theoretic (combinatorial) approach. In general, we will refer to simple graphs $G = (V, E)$ in which there are no self-loops or multiple-links. When directionality or weights of the links are considered it is explicitly stated; otherwise, we will refer to undirected and unweighted graphs. The concept of network communicability briefly described in the Introduction of this work immediately invokes the concept of walks in networks. A walk of length $k$ is a sequence of (not necessarily distinct) nodes $v_0, v_1, \cdots, v_{k-1}, v_k$ such that for each $i = 1, 2 \cdots, k$ there is a link from $v_{i-1}$ to $v_i$ (Cvetković *et al.*, 1997). Using the concept of walk we define the communicability between two nodes as follows. The ***communicability*** between the nodes $p$ and $q$ in a network is the weighted sum of all walks starting at node $p$ and ending at node $q$, in which the weighting scheme gives more weight to the shortest walks than to the longer ones.

Mathematically, the communicability function can be expressed as follows (Estrada and Hatano, 2008):

$$G_{pq} = \sum_{k=0}^{\infty} c_k \left( A^k \right)_{pq}, \tag{6}$$

where **A** is the adjacency matrix, which has unity in the $(p, q)$-entry if the nodes $p$ and $q$ are linked to each other and has zero otherwise. In Eq. (6), we have used the fact that the $(p, q)$-entry of the $k$th power of the adjacency matrix, $\left( A^k \right)_{pq}$, gives the number of walks of length $k$ starting at the node $p$ and ending at the node $q$ (Harary and Schwenck, 1979). The terms $G_{pp} = \sum_{k=0}^{\infty} c_k \left( A^k \right)_{pp}$ represent the *self-communicability* of a node and they provide a centrality measure known as the node *subgraph centrality* (Estrada and Rodríguez-Velazquez, 2005a). Centrality measures were originally introduced in social sciences (Freeman, 1979; Wasserman and Faust, 1994) and are now widely used in the whole field of complex network analysis (Newman, 2010). The coefficients $c_k$ need to fulfill the following requirements: (i) making the series (1) convergent, (ii) giving less weight to longer walks, and (iii) giving real positive values for the communicability. Then, assuming a factorial penalization we obtain the following communicability function:

$$G_{pq}^{EA} = \sum_{k=0}^{\infty} \frac{\left(A^k\right)_{pq}}{k!} = \left(e^A\right)_{pq}, \tag{7}$$

where $e^A$ is a matrix function that can be defined using the following Taylor series (Higham, 2008):

$$e^A = I + A + \frac{A^2}{2!} + \frac{A^3}{3!} + \cdots + \frac{A^k}{k!} + \cdots. \tag{8}$$

Note that the inclusion of the identity matrix in the expansion (8) does not affect neither the subgraph centrality nor the communicability between pairs of nodes since in the first case it only adds a constant to every value of the centrality measures and in the second case the off-diagonal entries are unchanged. Using the spectral decomposition of the adjacency matrix, the communicability function can be expressed as:

$$G_{pq}^{EA} = \sum_{j=1}^{n} \phi_{j,A}(p) \phi_{j,A}(q) e^{\lambda_{j,A}}, \tag{9}$$

where $\lambda_{1,A} \geq \lambda_{2,A} \geq \cdots \geq \lambda_{n,A}$ are the eigenvalues of the adjacency matrix in a non-increasing order and $\phi_{j,A}(p)$ is the $p$ th entry of the $j$ th eigenvector which is associated with the eigenvalue $\lambda_{j,A}$ of the adjacency matrix.

It is straightforward to realize that the shortest paths connecting any pair of nodes always make the largest contribution to the communicability function. That is, if $P_{rs}^{(l)}$ is the number of shortest paths between the nodes $r$ and $s$ having length $l$ and $W_{rs}^{(k)}$ is the number of walks of length $k > l$ connecting the same nodes, the communicability function is given by

$$G_{rs}^{EA} = \frac{P_{rs}^{(l)}}{l!} + \sum_{k>l} \frac{W_{rs}^{(k)}}{k!}, \qquad (10)$$

which indicates that $G_{pq}^{EA}$ accounts for all channels of communication between two nodes, giving more weight to the shortest path connecting them. Therefore, the name of '*communicability*' has been proposed to designate this function.

We can generalize the concept of communicability in three different ways. First, the analogy of the communicability with the thermal Green's function in statistical physics motivates us to introduce the temperature $T$, or its inverse $\beta$ as a weighting parameter:

$$G_{pq}^{EA} = \sum_{k=0}^{\infty} \frac{\left(\beta^k A^k\right)_{pq}}{k!} = \left(e^{\beta A}\right)_{pq}, \qquad (11)$$

where

$$e^{\beta A} = \beta I + \beta A + \frac{(\beta A)^2}{2!} + \frac{(\beta A)^3}{3!} + \cdots + \frac{(\beta A)^k}{k!} + \cdots. \qquad (12)$$

The 'physical meaning' of this parameter $\beta$ will be evident in the next sections of this paper.

An interesting way of utilizing the parameter $\beta$ is to consider the negative temperature. The communicability function of a network can be separated into the contributions coming from walks of even and odd lengths. For instance, for the case of $G_{pq}^{EA}$ we can write

$$\begin{aligned} G_{pq}^{EA} &= \sum_{j=1}^{n} \phi_{j,A}(p)\phi_{j,A}(q)\cosh(\lambda_{j,A}) + \sum_{j=1}^{n} \phi_{j,A}(p)\phi_{j,A}(q)\sinh(\lambda_{j,A}) \\ &= G_{pq}^{EA}(\text{even}) + G_{pq}^{EA}(\text{odd}) \end{aligned} \qquad (13)$$

We can separate these two contributions as

$$G_{pq}^{EA}(\text{even}) = \frac{1}{2}\left[G_{pq}^{EA}(\beta) + G_{pq}^{EA}(-\beta)\right],$$

$$G_{pq}^{EA}(\text{odd}) = \frac{1}{2}\left[G_{pq}^{EA}(\beta) - G_{pq}^{EA}(-\beta)\right].$$

For a network having link weights $w_{ij} \in \mathbb{R}^+$, the communicability function is obtained by using the weighted adjacency matrix $W = (w_{ij})_{n \times n}$ as

$$G_{pq}^{EA} = \sum_{k=0}^{\infty} \frac{(W^k)_{pq}}{k!} = (e^W)_{pq}. \tag{14}$$

In this case it has been proposed to normalize the weighted adjacency matrix in order to avoid the excessive influence of links with higher weights in the network (Crofts *et al.*, 2009). The normalization used so far transforms the weighted adjacency matrix as: $\tilde{W} = K^{-1/2}WK^{-1/2}$, where $K$ is a diagonal matrix of weighted degrees.

The third useful generalization of the communicability function is obtained by considering other penalization coefficients $c_k$ in the expression (6), which can give rise to different matrix functions. For instance, let $\alpha < 1/\lambda_1$ and let us take $c_k = \alpha^{-k}$ in Eq. (6) (Estrada and Higham, 2010). Then, we obtain the following communicability function:

$$G_{pq}^{RA} = \sum_{k=0}^{\infty} c_k(W^k)_{pq} = \sum_{k=0}^{\infty} \alpha^k(W^k)_{pq} = (I - \alpha W)^{-1}, \tag{15}$$

where we have replaced the adjacency matrix in (6) by its weighted version. The index $G_{pq}^{RA}$ was introduced as early as 1953 by Katz (1953) as a centrality measure for the nodes in social networks.

Another extension of the communicability function makes use of a strategy to increase or decrease the contribution of longer walks to the communicability between two nodes. For instance, indices zooming-in around a node give rise to the so-called $\psi_t(A)$ matrix functions (Estrada 2010a). In a similar way we can zoom out around a node by penalizing less the long walks from one node to another (Estrada 2010a).

The sum of the subgraph centralities for all nodes in the network represents a global index for the network (Estrada, 2000; Estrada and Rodríguez-Velazquez, 2005), which is nowadays known as the Estrada index of a network (de la Peña *et al.*, 2007; Deng *et al.*, 2009; Gutman *et al.*, 2010):

$$EE(G) = tr(e^A) = \sum_{j=1}^{n} e^{\lambda_{j,A}}. \qquad (16)$$

**B. Some combinatorial formulae**

Most of the combinatorial analysis of communicability in networks has been devoted to the so-called Estrada index, for which several bounds and analytic expressions have been proposed. The interested reader is referred to the recent reviews of Deng *et al.* (2009) and Gutman *et al.* (2010). Here we reproduce some expressions that can be useful for understanding these indices when analyzing complex networks. The Estrada index of a path or linear chain having $n$ nodes is given by (Gutman and Graovac, 2007)

$$EE(P_n) = \sum_{r=1}^{n} e^{2\cos(2r\pi/(n+1))}. \qquad (17)$$

Intuitively, the communicability between the two nodes at the end of a linear path should tend to zero as the length of the path tends to infinity. We can write the expression for $G_{rs}$ for the path $P_n$ (Estrada and Hatano, 2008):

$$G_{rs}^{EA}(P_n) = \frac{1}{n+1}\left(\sum_j \cos\frac{j\pi(r-s)}{n+1} - \cos\frac{j\pi(r+s)}{n+1}\right) e^{2\cos\left(\frac{j\pi}{n+1}\right)}. \tag{18}$$

Then it is straightforward to realize by simple substitution in (18) that $G_{rs}^{EA}(P_n) \to 0$ for the nodes at the end of a linear path as $n \to \infty$.

The Estrada index of a complete graph of $n$ nodes $K_n$, i.e., one having $n(n-1)/2$ links, is given by:

$$EE(K_n) = e^{n-1} + (n-1)e^{-1}, \tag{19}$$

and the communicability between any pair of nodes in the complete network $K_n$ is given by (Estrada and Hatano, 2008)

$$G_{rs}^{EA}(K_n) = \frac{e^{n+1}}{n} + e^{-1}\sum_{j=2}^{n}\phi_j(r)\phi_j(s) = \frac{e^{n+1}}{n} - \frac{1}{ne} = \frac{1}{ne}(e^n - 1). \tag{20}$$

This means that $G_{rs}^{EA} \to \infty$ as $n \to \infty$, which perfectly agrees with our intuition of what the communicability should mean in a network. For an Erdös-Rényi random graph with $n$ nodes and probability $p$, $G_{n,p}$ Shang (2011a) has found that the Estrada index is given by:

$$EE(G_{n,p}) = [1 + o(1)]e^{np}, \tag{21}$$

almost surely, as $n \to \infty$.

In a regular graph with $n$ nodes of degree $d = q+1$, the mean Estrada index $EE_{mean}(G,\beta) = EE(G,\beta)/n$ was found by Ejov et al. (2007) to be

$$EE_{mean}(G,\beta) = \frac{q+1}{2\pi}\int_{-2\sqrt{q}}^{2\sqrt{q}} e^{\beta s}\frac{\sqrt{4q-s^2}}{(q+1)^2 - s^2}ds + \frac{1}{n}\sum_{\gamma}\sum_{k=1}^{\infty}\frac{l(\gamma)}{2^{kl(\gamma)/2}}I_{kl(\gamma)}(2\sqrt{q}\beta), \tag{22}$$

where $\gamma$ runs over all (oriented) primitive geodesics in the network, $l(\gamma)$ is the length of $\gamma$, and $I_m(z)$ is the Bessel function of the first kind

$$I_m(z) = \sum_{r=0}^{\infty} \frac{(z/2)^{n+2r}}{r!(n+r)!}. \tag{23}$$

These authors (Ejov *et al.*, 2007) have observed a pattern of self-similarity named by them as *filars* when the average of the Estrada index is plotted against the variance of the same index. As displayed (Ejov *et al.*, 2007) for cubic graphs, the mean-variance plot form thread-like clusters with similar slopes and distances between consecutive clusters. It was shown that the graphs belonging to each cluster have the same number of triangles, and these numbers strictly increase from the left-most cluster to the right-most, starting from zero. Consequently, the mean-variance plot for regular graphs constitutes a way of characterizing the structure of these kinds of graph. Ejov *et al.* (2009) have demonstrated that this self-similar pattern is also observed for the mean-variance plot of the resolvent-like version of the Estrada index, which is derived from Eq. (12).

## III. PHYSICAL ANALOGIES
## A. Oscillator Networks

In the Introduction, we emphasized the analogy between the concept of correlation in physical systems and the communicability in network sciences. Here we explore the analogy more precisely, relating abstract complex networks with a physical oscillator model. In the present section, we consider every node as a ball of mass $m$ and every link as a spring with the spring constant $m\omega^2$ connecting two balls. We consider that the ball-spring network is submerged into a thermal bath at the temperature $T$. Then the balls in the complex network oscillate under thermal disturbances. How do the thermal disturbances propagate through the network? This physical analogy indeed gives the communicability of the complex network. For the sake of simplicity, we assume that there is no damping and no external forces are applied to the system. The coordinates chosen to describe a configuration of the system are $x_i$, $i = 1, 2, \cdots, n$, each of which indicates the fluctuation of the ball $i$ from its equilibrium point $x_i = 0$. Similar models have been previously used by Kim *et al.* (2003), who introduced the term *netons* to refer to phonons in a complex network in order to differentiate the underlying topological structure of these systems, which is not the usual periodic lattice.

## B. Network Hamiltonians

Let us start with a Hamiltonian of the oscillator network of the form

$$H_A = \sum_i \left[ \frac{p_i^2}{2m} + (K - k_i) \frac{m\omega^2 x_i^2}{2} \right] + \frac{m\omega^2}{2} \sum_{\substack{i,j \\ (i<j)}} A_{ij} (x_i - x_j)^2, \tag{24}$$

where $k_i$ is the degree of the node $i$ (the number of links that are connected to the node $i$) and $K$ is a constant satisfying $K \geq \max_i k_i$. The second term of the right-hand side is the potential energy of the springs connecting the balls, because $x_i - x_j$ is the extension or the contraction of the spring connecting the nodes $i$ and $j$. The first term in the first set of square parentheses is the kinetic energy of the ball $i$, whereas the second term in the first set of square parentheses is a counter term that offsets the movement of the network as a whole by tying the network to the ground. We add this term because we are only interested in small oscillations around the equilibrium; this will be explained below again.

The Hamiltonian (24) is expanded as follows:

$$H_A = \sum_i \left[ \frac{p_i^2}{2m} + (K - k_i) \frac{m\omega^2 x_i^2}{2} \right]$$
$$+ \frac{m\omega^2}{2} \left[ \sum_{\substack{i,j \\ (i<j)}} \left( A_{ij} x_i^2 \right) + \sum_{\substack{i,j \\ (i<j)}} \left( A_{ij} x_j^2 \right) - 2 \sum_{\substack{i,j \\ (i<j)}} \left( A_{ij} x_i x_j \right) \right]. \tag{25}$$

We can rewrite the first and second terms in the second set of square parentheses as

$$\sum_{\substack{i,j \\ (i \ne j)}} A_{ij} x_i^2 = \sum_i k_i x_i^2, \tag{26}$$

while the third term can be rewritten as

$$-\sum_{\substack{i,j \\ (i \ne j)}} A_{ij} x_i x_j = -\sum_{i,j} x_i A_{ij} x_j. \tag{27}$$

Therefore, the final form of Eq. (25) is given by

$$H_A = \sum_i \left( \frac{p_i^2}{2m} + \frac{Km\omega^2}{2} x_i^2 \right) - \frac{m\omega^2}{2} \sum_{i,j} x_i A_{ij} x_j. \tag{28}$$

Note that the term (26) cancels the $k_i$-dependent part of the counter term in Eq. (24).

Let us next consider the Hamiltonian of the oscillator network in the form

$$H_L = \sum_i \frac{p_i^2}{2m} + \frac{m\omega^2}{2} A_{ij} \left( x_i - x_j \right)^2 \tag{29}$$

instead of the Hamiltonian $H_A$ in Eq. (24). Because the Hamiltonian $H_L$ lacks the springs that tie the whole network to the ground (the second term in the first set of parentheses in the right-hand side of Eq. (24)), this network can undesirably move as a whole. We will deal with this motion shortly.

The expansion of the Hamiltonian (29) as in Eqs. (25)-(28) now gives

$$H_L = \sum_i \left( \frac{p_i^2}{2m} + \frac{m\omega^2}{2} k_i x_i^2 \right) - \frac{m\omega^2}{2} \sum_{i,j} x_i A_{ij} x_j$$
$$= \sum_i \frac{p_i^2}{2m} + \frac{m\omega^2}{2} \sum_{i,j} x_i L_{ij} x_j, \tag{30}$$

where $L_{ij}$ denotes an element of the network Laplacian $L$. The network Laplacian is given by $L = D - A$, where $D$ is a diagonal matrix with $D_{ii} = k_i$, and is often used in analyzing diffusion phenomena on complex networks. That is why we referred to Eq. (30) as $H_L$.

**C. Network of Quantum Oscillators**

We start by considering the quantum-mechanical counterpart of the Hamiltonian $H_A$ in Eq. (24). In this case the momenta $p_j$ and the coordinates $x_i$ are not independent variables but they are operators that satisfy the commutation relation,

$$\left[ x_i, p_j \right] = i\hbar \delta_{ij}. \tag{31}$$

We use the boson creation and annihilation operators defined by

$$a_i^\dagger = \frac{1}{\sqrt{2\hbar}} \left( x_i \sqrt{m\Omega} - \frac{i}{\sqrt{m\Omega}} p_i \right), \tag{32}$$

$$a_i = \frac{1}{\sqrt{2\hbar}} \left( x_i \sqrt{m\Omega} + \frac{i}{\sqrt{m\Omega}} p_i \right), \tag{33}$$

or

$$x_i = \sqrt{\frac{\hbar}{2m\Omega}} \left( a_i^\dagger + a_i \right), \tag{34}$$

$$p_i = \sqrt{\frac{\hbar}{2m\Omega}} \left( a_i^\dagger - a_i \right), \tag{35}$$

where $\Omega = \sqrt{K/m\omega}$. The commutation relation (31) yields

$$\left[a_i, a_j^\dagger\right] = \delta_{ij}. \tag{36}$$

With the use of these operators, we can recast the Hamiltonian (24), or equivalently Eq. (28), into the form

$$H_A = \sum_i \hbar\Omega\left(a_i^\dagger a_i + \frac{1}{2}\right) - \frac{\hbar\omega^2}{4\Omega}\sum_{i,j}\left(a_i^\dagger + a_i\right)A_{ij}\left(a_j^\dagger + a_j\right). \tag{37}$$

Since $A$ is symmetric, we can diagonalize it by means of an orthogonal matrix $O$ as in

$$\Lambda = O(KI - A)O^T, \tag{38}$$

where $\Lambda$ is the diagonal matrix with the eigenvalues $\lambda_\mu$ of $(KI - A)$ on the diagonal. This generates a new set of boson creation and annihilation operators as

$$b_\mu = \sum_i O_{\mu i} a_i = \sum_i a_i \left(O^T\right)_{i\mu}, \tag{39}$$

$$b_\mu^\dagger = \sum_i O_{\mu i} a_i^\dagger = \sum_i a_i^\dagger \left(O^T\right)_{i\mu}, \tag{40}$$

or

$$a_i = \sum_\mu \left(O^T\right)_{i\mu} b_\mu = \sum_\mu b_\mu O_{\mu i}, \tag{41}$$

$$a_i^\dagger = \sum_\mu \left(O^T\right)_{i\mu} b_\mu^\dagger = \sum_\mu b_\mu^\dagger O_{\mu i}. \tag{42}$$

Applying the transformations (39)-(42) to the Hamiltonian (37), we can decouple it as

$$H_A = \sum_\mu H_\mu, \tag{43}$$

with

$$H_\mu \equiv \hbar\Omega\left(b_\mu^\dagger b_\mu + \frac{1}{2}\right) + \frac{\hbar\omega^2}{4\Omega}(\lambda_\mu - K)(b_\mu^\dagger + b_\mu)^2$$

$$= \hbar\Omega\left(b_\mu^\dagger b_\mu + \frac{1}{2}\right) + \frac{\hbar\omega^2}{4\Omega}(\lambda_\mu - K)\left[(b_\mu^\dagger)^2 + (b_\mu)^2 + b_\mu^\dagger b_\mu + b_\mu b_\mu^\dagger\right] \qquad (44)$$

$$= \hbar\Omega\left[1 + \frac{\omega^2}{2\Omega^2}(\lambda_\mu - K)\right]\left(b_\mu^\dagger b_\mu + \frac{1}{2}\right) + \frac{\hbar\omega^2}{4\Omega}(\lambda_\mu - K)\left[(b_\mu^\dagger)^2 + (b_\mu)^2\right].$$

In order to go further, we now introduce an approximation in which each mode of oscillation does not get excited beyond the first excited state. In other words, we restrict ourselves to the space spanned by the ground state (the vacuum) $|\text{vac}\rangle$ and the first excited states $b_\mu^\dagger |\text{vac}\rangle$. Then the second term in the last line of the Hamiltonian (44) does not contribute and we thereby have

$$H_\mu = \hbar\Omega\left[1 + \frac{\omega^2}{2\Omega^2}(\lambda_\mu - K)\right]\left(b_\mu^\dagger b_\mu + \frac{1}{2}\right) \qquad (45)$$

within this approximation. This approximation is justified when the energy level spacing $\hbar\Omega$ is much greater than the energy scale of external disturbances, (specifically the temperature fluctuation $k_B T = 1/\beta$, in assuming the physical metaphor that the complex network is submerged into a thermal bath at the temperature $T$), as well as than the energy of the network springs $\hbar\omega$, i.e. $\beta\hbar\Omega \gg 1$ and $\Omega \gg \omega$. This happens when the mass of each oscillator is small, when the springs to the ground, $m\Omega^2$, are strong, and when the network springs $m\omega^2$ are weak. Then an oscillation of tiny amplitude propagates over the network. We are going to work in this limit hereafter. The thermal bath represents here an 'external situation' which affects all the links in the network at the same time, e.g., economic crisis, social agitation, extreme physiological conditions, etc. After equilibration, all links in the network are weighted by the parameter $\beta = (k_B T)^{-1}$. The parameter $\beta$ is known as the *inverse temperature* and $k_B$ is the Boltzmann constant. This is exactly the same parameter as the one that we have introduced in the previous section as a weight for every link in the network.

We are now in a position to compute the partition function as well as the thermal Green's function quantum-mechanically. As stated above, we consider only the ground state and one excitation from it. Therefore we have the quantum-mechanical partition function in the form

$$\begin{aligned} Z^A &= \langle \text{vac} | e^{-\beta H_A} | \text{vac} \rangle \\ &= \prod_\mu \langle \text{vac} | e^{-\beta H_\mu} | \text{vac} \rangle \\ &= \prod_\mu \exp\left\{ -\frac{\beta \hbar \Omega}{2} \left[ 1 + \frac{\omega^2}{2\Omega^2} (\lambda_\mu - K) \right] \right\}. \end{aligned} \qquad (46)$$

The diagonal thermal Green's function is given in the framework of quantum mechanics by

$$G_{pp}^A(\beta) = \frac{1}{Z} \langle \text{vac} | a_p e^{-\beta H_A} a_p^\dagger | \text{vac} \rangle, \qquad (47)$$

which indicates how much an excitation at the node $p$ propagates throughout the network before coming back to the same node and being annihilated. The transformations (39)-(42) let us compute the quantity (47) as

$$G_{pp}^A(\beta) = \frac{1}{Z^A}\sum_{\mu,\nu}(O^T)_{p\mu}\langle\text{vac}|b_\mu e^{-\beta H_A}b_\nu^\dagger|\text{vac}\rangle O_{\nu p}$$

$$= \frac{1}{Z^A}\sum_{\mu}\left[(O^T)_{p\mu}\langle\text{vac}|b_\mu e^{-\beta H_\mu}b_\mu^\dagger|\text{vac}\rangle O_{\mu p}\prod_{\nu(\neq\mu)}\langle\text{vac}|e^{-\beta H_\nu}|\text{vac}\rangle\right]$$

$$= \sum_{\mu}(O^T)_{p\mu}\frac{\langle\text{vac}|b_\mu e^{-\beta H_\mu}b_\mu^\dagger|\text{vac}\rangle}{\langle\text{vac}|e^{-\beta H_\mu}|\text{vac}\rangle}O_{\mu p} \qquad (48)$$

$$= \sum_{\mu}(O^T)_{p\mu}\exp\left\{-\beta\hbar\Omega\left[1+\frac{\omega^2}{2\Omega^2}(\lambda_\mu-K)\right]\right\}O_{\mu p}$$

$$= e^{-\beta\hbar\Omega}\left(\exp\left[\frac{\beta\hbar\omega^2}{2\Omega}A\right]\right)_{pp},$$

where we have used Eq. (38) in the last line. Similarly, we can compute the off-diagonal thermal Green's function as

$$G_{pq}^A(\beta) = e^{-\beta\hbar\Omega}\left(\exp\left[\frac{\beta\hbar\omega^2}{2\Omega}A\right]\right)_{pq}. \qquad (49)$$

Then, if we compare Eq. (49) with Eq. (7) we see that

$$G_{pq}^{EA} = e^{\beta\hbar\Omega}G_{pq}^A(\beta)$$

with the identification $\beta\hbar\omega^2 = 2\Omega$. Note that the constant $K$ affects only the proportionality constant through $\Omega = \sqrt{K/m\omega}$ in the expression (49). This means that when the temperature tends to infinite, $\beta \to 0$, there is absolutely no communicability between any pair of nodes. That is, $G_{pq}^{EA}(\beta \to 0) = (e^0)_{pq} = I_{pq} = 0$. An analogous situation to consider is that there is no way for the information to go from one node to another when all links in the network have been suppressed. If we consider the case when the temperature tends to zero, $\beta \to \infty$, then there is an infinite communicability between every pair of nodes, i.e., $G_{pq}^A(\beta \to \infty) = (e^\infty)_{pq} = \infty$.

The same quantum-mechanical calculation by using the Hamiltonian $H_L$ in Eq. (29) (instead of the Hamiltonian $H_A$ in Eq. (24)) gives

$$G_{pq}^L(\beta) = \lim_{\Omega \to 0} \left( \exp\left[ -\frac{\beta \hbar \omega^2}{2\Omega} L \right] \right)_{pq}$$
$$= 1 + \lim_{\Omega \to 0} O_{2p} O_{2q} \exp\left[ -\frac{\beta \hbar \omega^2}{2\Omega} \mu_2 \right], \quad (50)$$

where $\mu_2$ is the second eigenvalue of the Laplacian matrix. This gives the communicability function $G_{pq}^{EA}(\beta) + 1$ upon setting $\beta \hbar \omega^2 = 2\Omega$. Obviously, the term +1 added to the communicability function in the previous line does not have any effect for the practical use of this network measure. The reason why the second eigenvalue emerges in Eq. (50) is because the Laplacian matrix of a connected network has a zero eigenvalue as its first eigenvalue. This zero eigenvalue represents the mode where all nodes move in the same direction. Because the network represented by $H_L$ is not tied to the ground, nothing prevents the movement of the network as a whole. Since we are only interested in small oscillations around the equilibrium, we remove the mode of the zero eigenvalue from the above consideration and hence have the second eigenvalue in Eq. (50) as the first non-trivial eigenvalue.

In closing we have that:

> The communicability functions $G_{pq}^{EA}$ and $G_{pq}^{EL}$ of a network correspond to the thermal Green's function of a network of quantum harmonic oscillators.

## D. Network of Classical Oscillators

Let us now consider the classical-mechanical version of the Hamiltonian $H_A$ in Eq. (24). In classical mechanics, the momenta $p_i$ and the coordinates $x_i$ are independent variables. In statistical mechanics of classical systems, the integration of the factor

$$\prod \exp\left[-\beta\left(\frac{p_i^2}{2m}\right)\right] \tag{51}$$

over the momenta $\{p_i\}$ reduces to a constant term, not affecting the integration over $\{x_i\}$. We will therefore leave out the kinetic energy for the moment and consider the Hamiltonian of the form

$$\begin{aligned} H_A &= \frac{Km\omega^2}{2}\sum_i x_i^2 - \frac{m\omega^2}{2}\sum_{i,j} x_i A_{ij} x_j \\ &= \frac{m\omega^2}{2} x^T (KI - A) x, \end{aligned} \tag{52}$$

where $x = (x_1, x_2, \cdots, x_n)^T$ and $I$ is the $n \times n$ identity matrix.

Let us calculate the partition function $Z$ and the thermal Green's function $G_{pq}$ in the framework of classical statistical mechanics. The partition function is given by

$$Z = \int e^{-\beta H_A} \prod_i dx_i = \int dx \exp\left(-\frac{\beta m\omega^2}{2} x^T (KI - A) x\right), \tag{53}$$

where the integral is $n$-fold. We can carry out this $n$-fold integration by diagonalizing the matrix $A$. Now, we can use the same diagonalization as in Eq. (28). By taking a sufficiently large value of the constant $K$, we can make all eigenvalues $\lambda_\mu$ positive. By defining a new set of variables $y_\mu$ as $y = Ox$ and $x = O^T y$, we can transform the Hamiltonian (52) in the form

$$H_A = \frac{m\omega^2}{2} y^T \Lambda y = \frac{m\omega_0^2}{2}\sum_\mu y_\mu^2 + \frac{m\omega^2}{2}\sum_\mu \lambda_\mu y_\mu^2. \tag{54}$$

On the other hand, the integration measure of the $n$-fold integration in Eq. (53) is transformed as $\prod_i dx_i = \prod_\mu dy_\mu$, because the Jacobian of the orthogonal matrix $O$ is unity. Therefore, the multi-fold integration in the partition function (53) is decoupled to give

$$Z = \prod_\mu \left[ \int \exp\left(-\frac{\beta m\omega^2}{2}\lambda_\mu y_\mu^2\right) dy_\mu \right] \tag{55}$$

$$= \prod_\mu \sqrt{\frac{2\pi}{\beta m\omega^2 \lambda_\mu}}. \tag{56}$$

We can rewrite this in terms of the original matrix $A$ in the form

$$Z = \left(\frac{2\pi}{\beta m}\omega^2\right)^{n/2} \frac{1}{\sqrt{\det(KI - A)}}. \tag{57}$$

Since we have made all the eigenvalues of $(KI - A)$ positive, its determinant is positive.

The centrality index may be given in the framework of classical mechanics by

$$G_{pp}(\beta) = \langle x_p^2 \rangle = \frac{1}{Z}\int x_p^2 e^{-\beta H_A} \prod_i dx_i. \tag{58}$$

The same transformation as in Eqs. (54)-(58) yields

$$G_{pp}(\beta) = \frac{1}{Z}\int \left[\sum_\sigma \left(O^T\right)_{p\sigma} y_\sigma\right]^2 e^{-\beta H_A} \prod_\mu dy_\mu. \tag{59}$$

In the integrand, odd functions with respect to $y_\mu$ vanish. Therefore, only the terms of $y_\sigma^2$ survive after integration in the expansion of the square parentheses in the integrand. This gives

$$G_{pp}(\beta) = \frac{1}{Z} \int \left[ \sum_\sigma (O_{\sigma p} y_\sigma)^2 \right] \exp\left( -\frac{\beta m \omega^2}{2} \sum_\nu \lambda_\nu y_\nu^2 \right) \prod_\mu dy_\mu$$

$$= \frac{1}{Z} \sum_\sigma O_{\sigma p}^2 \int y_\sigma^2 \exp\left( -\frac{\beta m \omega^2}{2} \lambda_\sigma y_\sigma^2 \right) dy_\sigma \qquad (60)$$

$$\times \prod_{\mu(\neq \sigma)} \left[ \int \exp\left( -\frac{\beta m \omega^2}{2} \lambda_\mu y_\mu^2 \right) dy_\mu \right].$$

Comparing this expression with Eq. (55), we have

$$G_{pp}(\beta) = \sum_\sigma O_{\sigma p}^2 \left( \frac{\int y_\sigma^2 e^{-\beta m \omega^2 \lambda_\sigma y_\sigma^2 / 2} dy_\sigma}{\int e^{-\beta m \omega^2 \lambda_\sigma y_\sigma^2 / 2} dy_\sigma} \right) = \sum_\sigma O_{\sigma p}^2 \frac{\sqrt{\left[\frac{2\pi}{\beta m \omega^2 \lambda_\sigma}\right]^3}}{\sqrt{\frac{2\pi}{\beta m \omega^2 \lambda_\sigma}}}$$

$$= \sum_\alpha \frac{O_{\sigma p}^2}{\beta m \omega^2 \lambda_\sigma} \qquad (61)$$

$$= \frac{1}{\beta m \omega^2} \left[ (KI - A)^{-1} \right]_{pp}$$

$$= \frac{1}{\beta m K \omega^2} \left[ (I - A/K)^{-1} \right]_{pp}.$$

Likewise, the communicability measure may be given by the thermal Green's function in the framework of classical mechanics as

$$G_{pq}(\beta) = \langle x_p x_q \rangle = \frac{1}{Z} \int x_p x_q e^{-\beta H_A} \prod_i dx_i,, \qquad (62)$$

which results in

$$G_{pq}(\beta) = \frac{1}{\beta K m \omega^2} \left[ (I - A/K)^{-1} \right]_{pq}. \tag{63}$$

This represents a correlation between the node displacements in a network due to small thermal oscillations (Estrada and Hatano, 2010a; b). Comparing the last expression with Eq. (21), we arrive at

$$G_{pq}^{RA} = \beta m K \omega^2 G_{pq}(\beta)$$

with the identification $\alpha = 1/K$.

The same calculation using the Hamiltonian (29) gives

$$G_{pq}^{D}(\beta) = \frac{1}{\beta m \omega^2} (L^+)_{pq} \tag{64}$$

where $L^+$ is the Moore-Penrose generalized inverse of the Laplacian. This is due to the fact that the Laplacian matrix of a connected network has a nondegenerate zero eigenvalue. Based on similar considerations to those described below Eq. (50), we remove the mode of the zero eigenvalue from the above consideration and hence have (64).

Then, we conclude that:

> The communicability functions $G_{pq}^{RA}$ and $G_{pq}^{D}$ of a network correspond to the thermal Green's function of a network of classical harmonic oscillators.

## IV. COMPARING COMMUNICABILITY FUNCTIONS

In the previous sections we have defined four communicability functions, two based on networks of quantum harmonic oscillators $G_{pq}^{EA}$ and $G_{pq}^{EL}$, and two on networks of classical harmonic oscillators $G_{pq}^{RA}$ and $G_{pq}^{D}$. As was emphasized in Section I.C, there is not a systematic way of selecting one communicability function for a particular problem; the use of one or another of these functions relies on the particular problem under study. Consequently, we give here a couple of examples to illustrate the use of these communicability functions in different scenarios.

### A. Study of a social conflict

The first example consists of a small social network studied by Thurman (1979) as a result of 16 months of observation of office politics. The office was an overseas branch of a large international organization and consisted of 15 members. Thurman studied an informal network of friendship ties among the members, which was not a part of the official structure of the office. During Thurman's study a conflict arose in the office as two members, identified as Emma and Minna, were the targets of a leveling coalition formed by 6 members of the staff, identified as Ann, Amy, Katy, Pete, Tina, and Lisa. The attacking coalition is formed by some of the best connected members of the office. However, Emma, who is one of the targets of the attacks has as many connections as Pete and Ann, which are in the coalition. On the other hand, Minna has the same number of ties as Andy and Bill, which are not the objective of the coalition.

Let us consider the average communicability for a given node defined as

$$\langle G_p \rangle = \frac{1}{n-1} \sum_{q \neq p}^{n} G_{pq}.$$

In FIG. 1a we illustrate the social network of the overseas office in which the targets of the coalition are drawn in black and the members of the coalition in gray. In FIG. 1b we plot the values of the normalized average communicability for every individual in the office for the four different kinds of communicability previously defined. The quantum and classical communicabilities are in general linearly related to each other. For instance, the Pearson correlation coefficient between $\langle G_p^{EA} \rangle$ and $\langle G_p^{RA} \rangle$ is 0.97. In this example we have not systematically varied the value of the empirical parameter $\alpha$ in $G_{pq}^{RA} = \left[\left(I - \alpha A\right)^{-1}\right]_{pq}$, which is a necessary but time consuming part of the use of resolvent-like communicability. Here we have used $\alpha = 0.1$, which fulfills the condition $\alpha < 1/\lambda_1$.

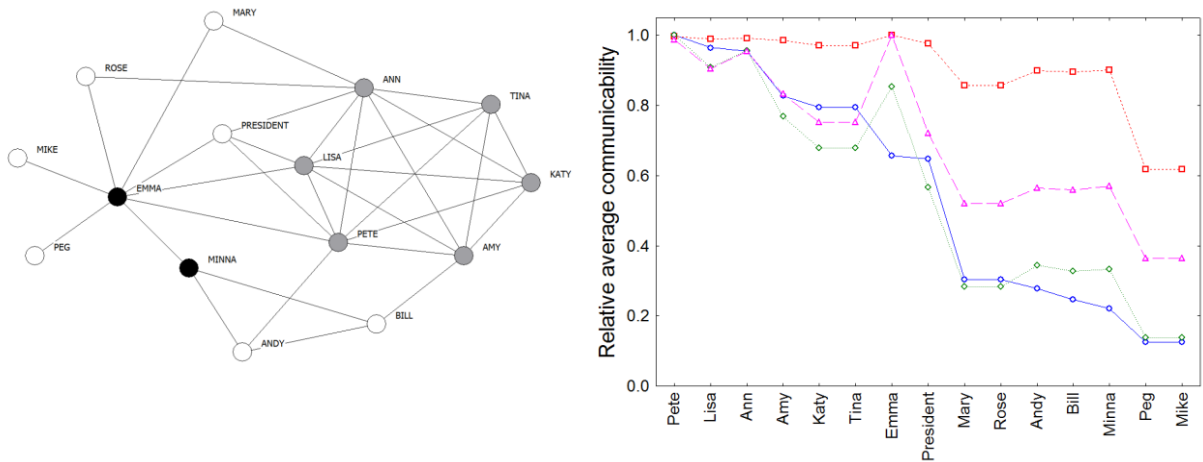

FIG. 1. (color online). a) Diagrammatic representation of the social network of friendship ties in the office of an overseas branch of a large international organization according to Thurman (1979). Members of the coalition are drawn in gray, and targets in black. b) Values of the relative communicability (see text) for every member of the social network represented in a). The values of the average communicabilities are as follows: squares $\left\langle G_p^{EL} \right\rangle$, circles $\left\langle G_p^{EA} \right\rangle$, diamonds $\left\langle G_p^{RA} \right\rangle$, triangles $1/\left\langle G_p^D \right\rangle$. Note that $\left\langle G_p^D \right\rangle$ increases as the other indices decrease, for which we have plotted $1/\left\langle G_p^D \right\rangle$ instead.

The main difference between the four communicability measures is that $\left\langle G_p^{EA} \right\rangle$ ($\beta \equiv 1$) is the only one that ranks the six members of the coalition as the ones having the largest average communicability among all members of the office (FIG. 1b). The highest communicability is observed between Pete and Lisa, Pete and Ann as well as Ann and Lisa. Pete has been recognized by Thurman as the center of the social circle in the office, which also involves Lisa, Katy and Amy. Pete was coming to the office from the central office and was known to have dated both Katy and Amy. He was also the one who arranged for Ann to be assigned to this office. Emma, who is one of the targets of the attacks, occupies the position immediately below the coalition and displays a good communicability with the President. It was known that Emma played an important role in the office as she was promoted to the administrative manager, where she has direct control of the drivers, the bookkeeping section, the secretarial pool and a variety of other services. Then, it is plausible that the members of the coalition see Emma as a threat, making her the target of their attacks. On the other hand, her relatively large communicability with all members of the organization makes her less vulnerable to the attacks of the coalition (see FIG. 2a). At the end of the day she was able to resist the attacks and consolidate her position in the office; as Thurman has put the case, "*she could mobilize a defense against a leveling coalition though a counter-attack*".

The situation of Minna was quite different. She is placed by the average communicability index $\left\langle G_p^{EA} \right\rangle$ at the bottom of the ranking together with Peg and Mike. She was new at the office as she came from another field office. Despite that she had over 20 years of experience, Pete and the president had been warned of Minna's "over enthusiasm", which might interrupt the smooth working of the office. Her lack of communicability makes her a very vulnerable target of the attacks. She was very much affected by this situation as "*she could not use her reticulum to mobilize effective support in a conflict situation*".

The analysis of this social conflict is very much complemented by the study of the node displacement correlation $G_{pq}^D$ among members of the office. Let us consider that the office as a whole has been 'shaken' by the conflict that has arisen there. Every member of the office will be affected, having some displacements $G_{pp}^D$ from their equilibrium position. The most robust members will be less affected and they will display only small displacements in comparison with more vulnerable ones. The sign of the term $G_{pq}^{D}$ will tell us whether two members of the office are 'correlated' in their displacements or not. That is, if two members of the office have the same sign for $G_{pq}^{D}$ we can assume that they are responding in a coordinated way to the 'thermal oscillations' of the network as a whole. Then, we have seen that the members of the coalition not only display small values of $G_{pp}^{D}$ indicating their robust position in the office but also that their displacements are positively correlated (see FIG. 2b).

Emma again has the most robust position in the office according to her very low value of $G_{pp}^{D}$, which can explain why she was so resilient to the attacks. However, the most revealing thing is provided by the values of $G_{pq}^{D}$ between Emma and the members of the coalition. As can be seen in FIG. 2b Emma is anticorrelated with the members of the attacking coalition and Minna is anticorrelated with all other members of the office. Emma is positively correlated with the president and with some of the weakest members (according to their communicability) of the office. These results indicate that it is not only important to have a robust position in the office but also to be positively coordinated with the important members of the office to avoid possible attacks of leveling coalitions. The current analysis could provide some support to the arduous labor of sociologists in their field work, in particular for the quantitative analysis of the effects producing conflicts in social systems.

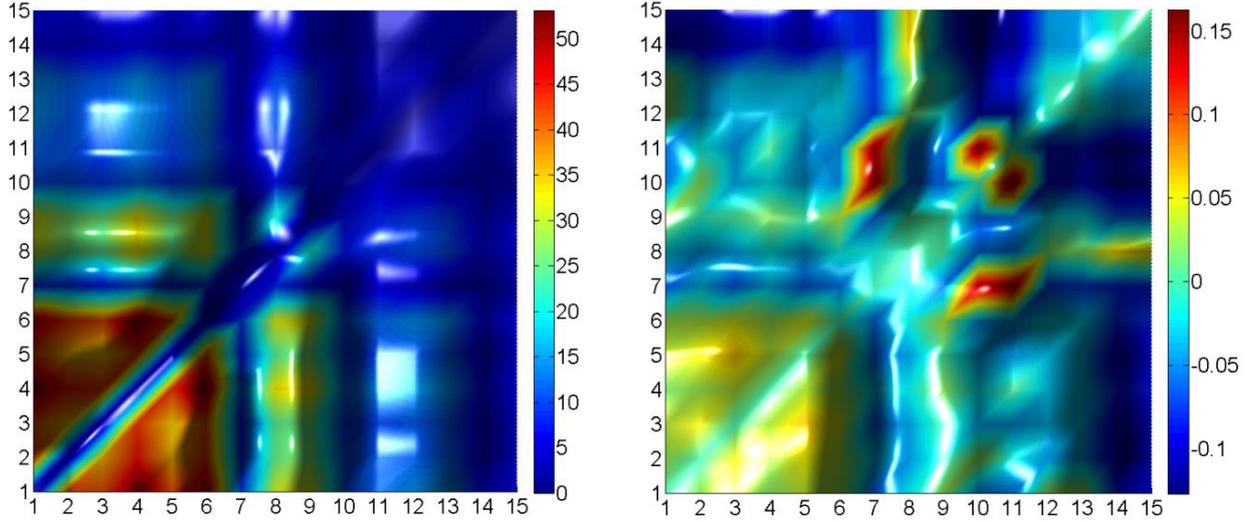

FIG. 2 (color online). Average communicability between members of the office according to $G_{pq}^{EA}$ (a) and to $G_{pq}^{D}$ (b). The members of the office are numbered as: 1: Ann, 2: Amy, 3: Katy, 4: Pete, 5: Tina, 6: Lisa, 7: Minna, 8: Emma, 9: President, 10: Bill, 11: Andy, 12: Mary, 13: Rose, 14: Mike, 15: Peg.

An important advantage of the consideration of the communicability based on a network of quantum harmonic oscillators is that we can explore the effect of the 'temperature' on the process under study. That is, while for the networks of classical oscillators the communicability changes linearly with the temperature (see Eqs. (63) and (64)), for the network of quantum oscillators the change of temperature affects non-trivially the structure of the network (see Eqs. (49) and (50)). Then, if we study $\left\langle G_p^{EA}(\beta) \right\rangle$ for the members of the overseas office we can observe some important changes that give important information about the evolution of the conflict in the office. For instance, as the temperature increases from $\beta = 1$ to $\beta = 0.5$ it can be seen that the gap in communicability between Emma and the leveling coalition decreases significantly as can be seen in FIG. 3. As the temperature increases to $\beta = 0.1$ Emma becomes one of the best communicated persons in the office only surpassed by Pete and Ann (see FIG. 3). The increase of temperature here can be understood as the increase of the tensions in the office and the relative increase of communicability of Emma can explain quantitatively the findings of Thurman that during the crisis Emma was able to consolidate her position in the office and even gaining more status.

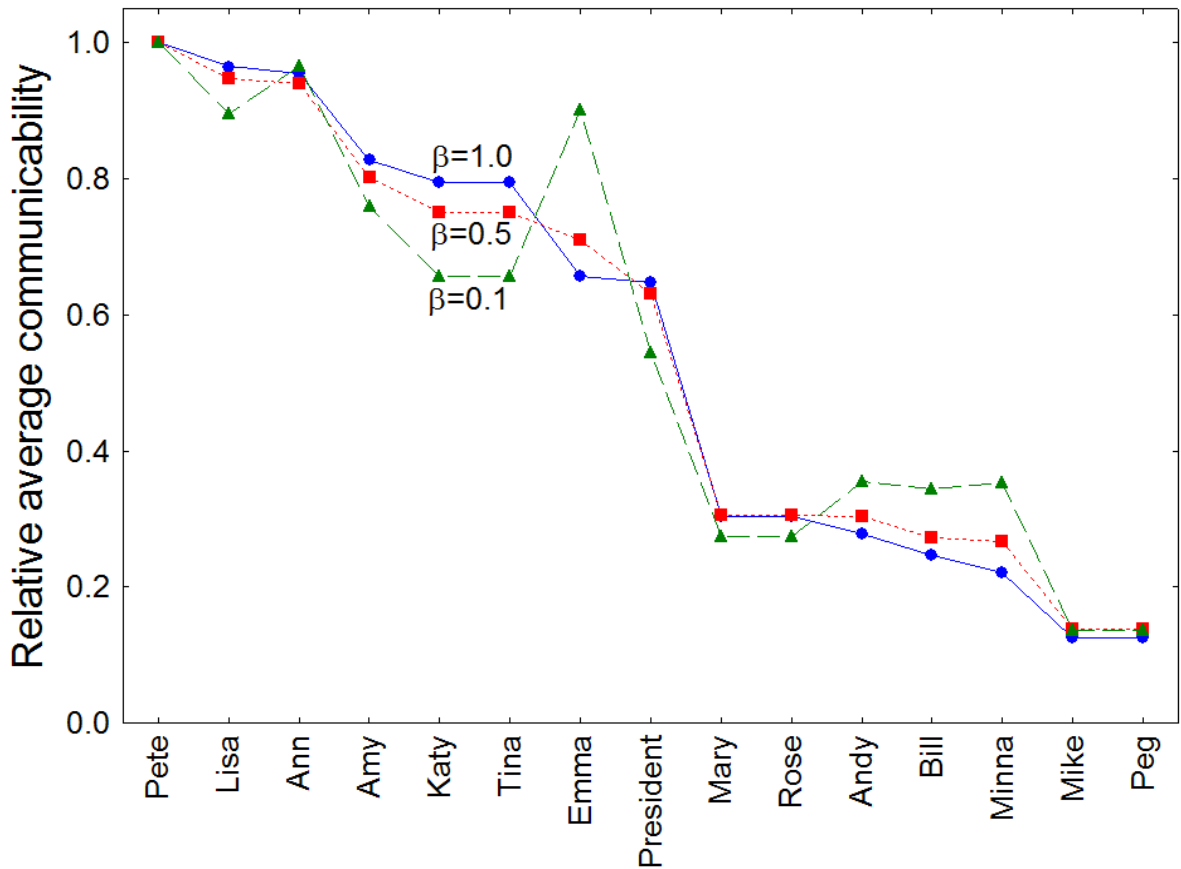

FIG 3 (color online). Relative average communicability for each member of the overseas office studied by Thurman (1979) at three different temperatures.

## B. Study of biomolecular networks

As a second example of the use of communicability functions in complex networks we illustrate the study of atomic motion in biomolecular systems. In this case there is an experimental measure that accounts for the displacement of atoms in such molecules due to the thermal oscillations. Such an experimental measure is provided by X-ray experiments as the so-called B-factor or the temperature factor, which represents the reduction of coherent scattering of X-rays due to the thermal motion of the atoms. The B-factors are very important for the study of protein structures as a measure of their dynamical behavior (Soheilifard *et al.*, 2008). For instance, regions with large B-factors are usually more flexible and functionally important. Bahar *et al.* (1997) used the atomic displacements $G_{pp}^{D}$ to describe thermal fluctuations in proteins.

Here we consider the protein lipase B from *Candida antarctica* (1tca) (Uppernberg, 1994) represented as a complex network in which the nodes represent amino acids, centred at their $C_{\beta}$ atoms, with the exception of glycine for which $C_{\alpha}$ is used. Two nodes are then connected if the distance $r_{ij}$ between both $C_{\beta}$ atoms of the residues $i$ and $j$ is not longer than a certain cutoff value $r_{C} = 7.0$ Å. In FIG. 4 we illustrate the values of the experimental B-factors (bottom) and those of the Laplacian-based atomic displacements obtained from consideration of the protein as a network of classical $G_{pp}^{D}$ (middle) and of quantum $G_{pp}^{EL}$ (top) harmonic oscillators (Estrada, 2010b). It can be seen that the atomic displacement $G_{pp}^{D}$ shows better correlation with the experimental values of the B-factor than $G_{pp}^{EL}$. However, in both cases, $G_{pp}^{D}$ and $G_{pp}^{EL}$, the region around the amino acid number 250 appears like the most flexible one, in contrast with the experimental results that show the region around the residue 220 as the one having the largest atomic displacements. Also the atomic displacements of the amino acids 70 and 124 appear exaggeratedly large (see FIG. 4). In fact, both communicability indices are linearly related with a Pearson correlation coefficient equal to 0.95.

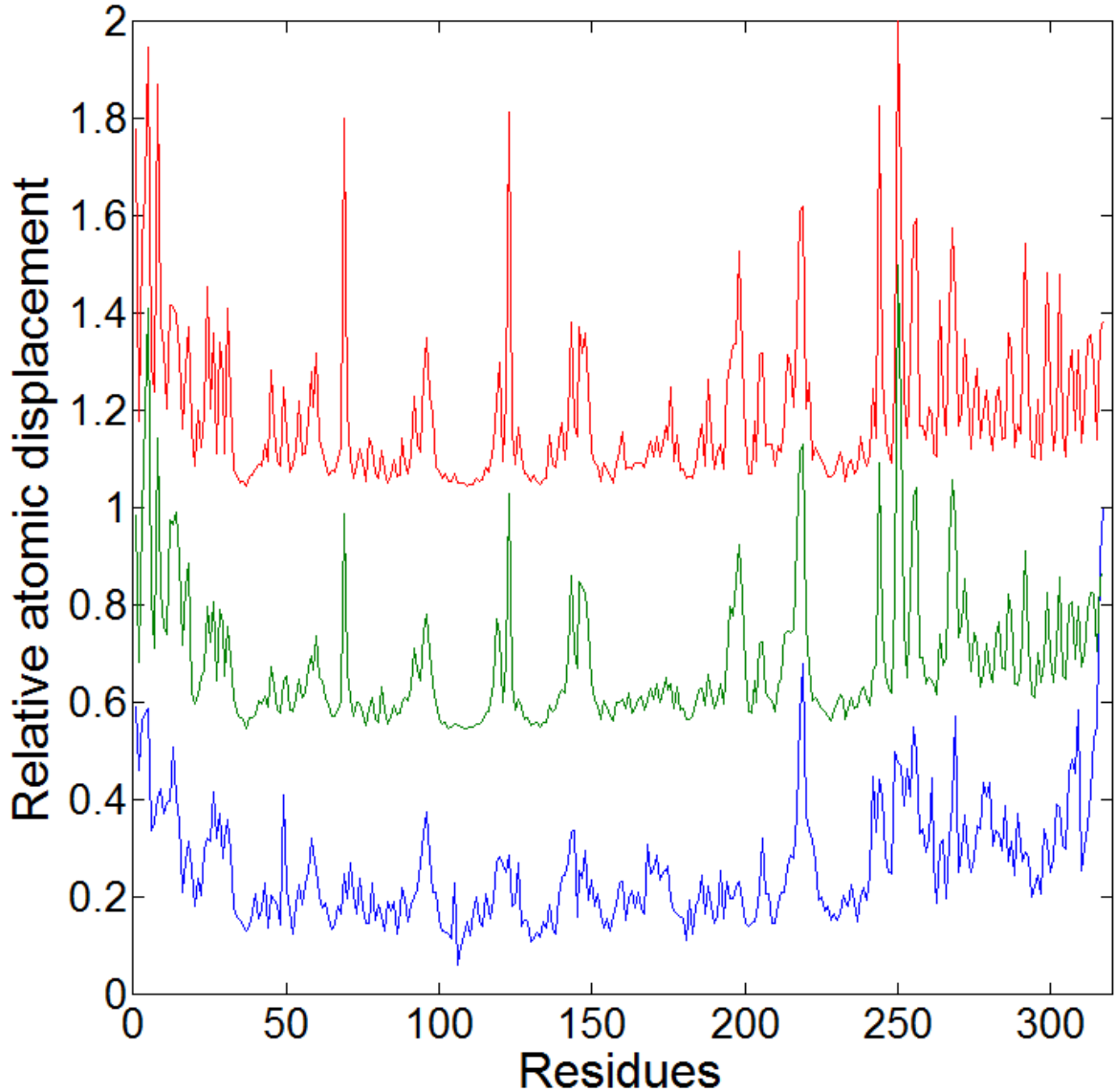

FIG. 4 (color online). Relative atomic displacement for the residues in the lipase B from *Candida antarctica* (1tca). Bottom curve: experimental B factors, middle curve: $G_{pp}^{D}$, top curve: $G_{pp}^{EL}$. The values of $G_{pp}^{D}$ and $G_{pp}^{EL}$ are displaced 0.5 and 1.0 units up in order to provide better visibility.

As we have seen in the analysis of the social network in the previous section the use of communicabilities based on networks of quantum harmonic oscillators gives the advantage of exploring the effects of the temperature on the processes under study. In FIG. 5 we plot the values of $G_{pp}^{EL}(\beta)$ for the residues in the protein 1tca at four different temperatures. As we can see, as soon as the temperature decreases the region around the amino acids 70, 124 and 250 start to lose their flexibility in comparison with that of the residue 220. At the same time the linear correlation coefficient between the experimental B-factors and the communicability increases from 0.66 for $\beta=1$ to 0.75 for $\beta=8$, which is even better than the value (0.71) obtained by using $G_{pp}^{D}$. For $\beta>8$, the relation between the experimental and the calculated values of B-factors becomes non-linear.

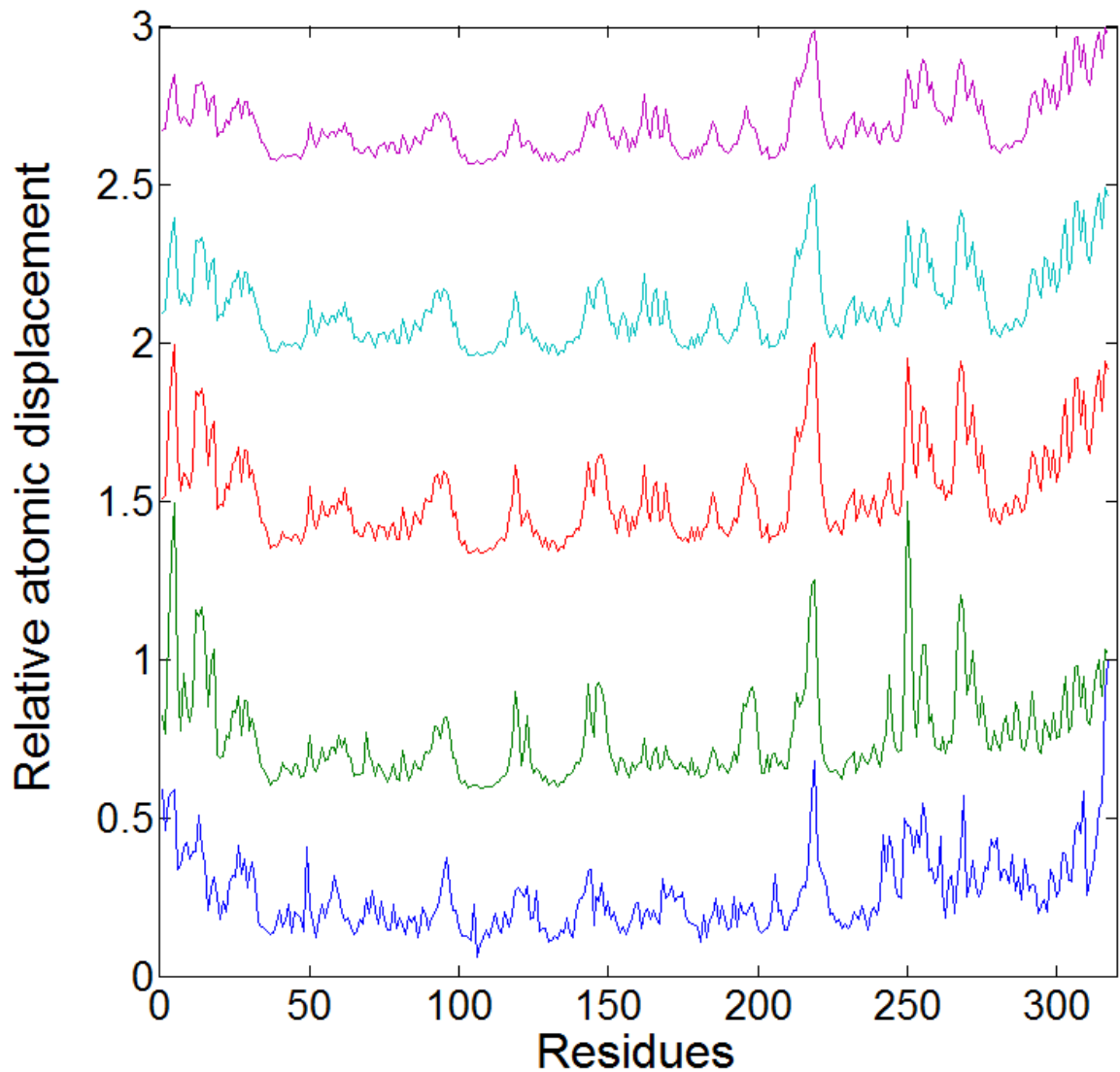

FIG. 5 (color online). Relative atomic displacement for the residues in the lipase B from *Candida antarctica* (1tca) at three different temperatures. From bottom to top: Exp., $\beta = 4$, $\beta = 8$, $\beta = 10$, $\beta = 12$. The values of the communicability are displaced up 0.5 units each in order to provide better visibility.

The two examples analyzed so far in this section, the social conflict and atomic displacements in proteins, have shown that in general communicabilities based on quantum and classical harmonic oscillators are linearly related to each other. This is repeated in many complex networks not analyzed in this review. As we have seen in these two examples, the non-trivial variation of the quantum-based communicabilities with the temperature and the necessity of using an empirical parameter for the classical one gives some advantages to the quantum communicability. However, there are situations, not analyzed so far in this review, in which the communicabilities based on classical oscillators are the appropriate choice. This is for instance the study of networks that evolve in time (Grindrod *et al.*, 2011). In this case the use of the classical communicability produces the right penalization of walks that evolve not only in one snapshot of a network, but also in a sequence of times. In the next section we will provide more general examples of the application of communicability functions for the analysis of a variety of processes in complex networks.

## V. COMMUNICABILITY AND THE ANALYSIS OF NETWORKS

We consider here the analysis of a complex network in three different scales: *micro-*, *meso-* and *macroscopic*. What we understand here as a '*microscopic*' analysis of a complex network is the consideration of its local topological properties, such as those derived from the analysis of close environments around individual nodes and links. An extension of this environment allows us to analyze a '*mesoscopic*' level of organization in which nodes and links group together forming some kind of clusters characterized by properties which are more or less independent of the properties of individual nodes and those of the network as a whole. The '*macroscopic*' properties of complex networks refer to their global topological properties. That is, those properties that characterizes the network as a whole. There have been several works analyzing the role of communicability and self-communicability at these three different scales (Bradonjić *et al.*, 2011; Crofts and Higham, 2009; Crofts *et al.*, 2011; da Fontoura Costa *et al.*, 2008; Došlić, 2005; Estrada 2007a; b; 2010b; Estrada and Bodin, 2008; Estrada and Hatano, 2009a; b; Estrada *et al.*, 2009; Jungsbluth *et al.*, 2007; Koponen and Pehkonen, 2010; Li *et al.*, 2010; Ma *et al.*, 2010; MacArthur *et al.*, 2010; Malliaros and Megalooikonomou, 2011; Ren *et al.*, 2011; Tordesillas *et al.*, 2010; Shang, 2011b; Wang and Qin, 2010; Walker and Tordesillas, 2010; Ying and Wu, 2008). We present here some illustrative examples to give a flavor of the relevance of these approaches.

**A. Microscopic analysis of networks**

An example of the use of communicability for analyzing the microscopic structure of networks is the identification of essential proteins in protein-protein interaction (PPI) networks. An essential protein is one that when knocked out renders the cell unviable. After a pioneering work of Jeong *et al.* (2001) a method was designed to identify essential protein *in silico* using the topological information provided by the PPI network (Estrada, 2006a, b). The method consists of ranking the proteins in the PPI network according to a given centrality measure. Then, it is expected that the top proteins in such ranking are essential to this organism. The *proof-of-concept* for this method was provided by the analysis of a small dataset of the yeast PPI network consisting of 2,224 proteins and 6,604 interactions (von Mering *et al.*, 2002). In this experiment the self-communicability (subgraph centrality) of a protein emerged as the best predictor for protein essentiality among 6 centrality measures (Estrada, 2006a). For instance, for the selection of the top 100 proteins $G_{pp}^{EA}$ identifies 54% of the essential proteins, while the degree identifies 43% and the random selection identifies 25%. More recent results have shown how to improve these percentages and more importantly how false positives affect the discovery of essential proteins by using centrality measures. Li *et al.* (2010) used three datasets with different levels of confidence which consists of 2,455, 11,000 and 45,000 interactions, respectively. They showed that lower percentages of essential proteins are identified by any centrality measures when the method is applied to less confidence datasets. Then, Li *et al.* (2010) used a strategy consisting of giving a weight to each interaction in the PPI network of yeast, which represents the probability of this interaction being a true positive. This confidence score for each interaction was assigned on the basis of two criteria: (1) observing experimental evidences for the interaction; (2) evaluating the function similarity of the pair of proteins using gene ontology (GO) semantic similarity. Li *et al.* (2010) studied a PPI network of yeast consisting of 4,746 proteins and 15,166 interactions and its high reliability core formed by 2,373 proteins and 5,283 interactions. In all the cases analyzed, the weighted subgraph centrality produced the best performance in identifying essential proteins. For instance, by selecting 10% in the total PPI network the weighted subgraph centrality identifies 53% of the existing essential proteins versus 44% identified by its non-weighted version. The second highest percentages are observed for the weighted versions of the information (Stephenson and Zelen, 1989) and eigenvector (Bonacich, 1972; 1987) centralities which identify 47% of essential proteins. When the experiment is carried out for the core of high reliability proteins, the weighted subgraph centrality identifies 55% of essential proteins versus 52% of its unweighted version.

One characteristic of PPI networks is that many proteins are grouped together in functional modules in which most of the proteins share some functionality (see further). Consequently, a microscopic analysis of the network is not enough for identifying essential proteins. For instance, if a protein has many links with other proteins which are in different modules, the knocking out of such protein will make many protein complexes disconnected with a consequent loss of different functionalities in the cell. This can be interpreted as a plausible cause for essentiality of that protein. Using this reasoning Ren *et al.* (2011) have modified the subgraph centrality to include the participation of a protein in protein complexes. This kind of approach can be considered as a combination of micro- and mesoscopic scales for the analysis of a network. They considered the number of links that a protein *i* has with other proteins in a complex $C$, $k(i,C)$. Then, the values of $k(i,C)$ are summed for all complexes in which the protein *i* takes place giving the number of links that the protein *i* has in different protein complexes, $mC_i$. The so-called harmonic centrality is now defined as:

$$HC_i = \frac{1}{2}\left(G_{ii}^{EA}/G_{max}^{EA} + mC_i/mC_{max}\right), \tag{65}$$

where the subscript '*max*' indicates the maximum value among all nodes of the network. The factor ½ in the expression was determined empirically. Using this approach Ren *et al.* (2011) studied two PPI networks from DIP database (Xenarios *et al.*, 2000), one in which protein complexes are identified by experimental methods (YGS_PC) and another in which they are identified by the CMC algorithm (YCMC_PC). The first includes 1,042 proteins and 209 complexes, while the second is formed by 1,538 proteins and 623 complexes. The results obtained by using this approach show that the $HC_i$ method reaches an impressive 70% of good classification for the top 200 ranked proteins.

**B. Mesoscopic analysis of networks**

Protein complexes are examples of a mesoscopic type of organization existing in complex networks. This organization consists of several clusters of tightly connected nodes forming distinguishable communities which are relatively poorly connected to each other. The detection of communities in complex networks has become one of the most intensive areas of interdisciplinary research in this field (Fortunato, 2010; Fortunato and Barthélemy, 2007; Newman, 2004; 2006a; b; von Luxburg, 2007). Estrada and Hatano (2008) have proposed to reveal the community structure of complex networks by using the sign separation of the communicability function:

$$G_{pq}^{EA} = \phi_{1,A}(p)\phi_{1,A}(q)e^{\lambda_{1,A}}$$
$$+ \left[ \sum_{2 \leq j \leq n}{}^{++} \phi_{j,A}(p)\phi_{j,A}(q)e^{\lambda_{j,A}} + \sum_{2 \leq j \leq n}{}^{--} \phi_{j,A}(p)\phi_{j,A}(q)e^{\lambda_{j,A}} \right] \quad (66)$$
$$+ \left[ \sum_{2 \leq j \leq n}{}^{+-} \phi_{j,A}(p)\phi_{j,A}(q)e^{\lambda_{j,A}} + \sum_{2 \leq j \leq n}{}^{-+} \phi_{j,A}(p)\phi_{j,A}(q)e^{\lambda_{j,A}} \right],$$

where $\sum^{++}$ represents the summation over the terms with both $\varphi_{j,A}(p)$ and $\varphi_{j,A}(q)$ positive, $\sum^{+-}$ represents the summation over the terms with both $\varphi_{j,A}(p)$ positive and $\varphi_{j,A}(q)$ negative, and so on.

In the vibrational approach in which the communicability is identified as the thermal Green's function of the network, the first term corresponds to the 'translational' movement of the network with all nodes vibrating in the same direction. The second term is identified with the coordinated vibrations of a pair of nodes which vibrates in the same direction. The third term corresponds to the 'discoordinate' vibration of the pairs of nodes in which one is moved in one direction and the other moves in the contrary one. Notice that the third term is negative. Therefore, the communicability function can be written as: $G_{pq}^{EA} = G_{pq}^{EA}(\text{tras}) + G_{pq}^{EA}(\text{coord}) - \left|G_{pq}^{EA}(\text{disc})\right|$. Then, we say that two nodes $p$ and $q$ are in the same cluster if their contribution to the communicability coming from $G_{pq}^{EA}(\text{coord})$ is larger than that coming from $\left|G_{pq}^{EA}(\text{disc})\right|$. In other words, two nodes are in the same community if they are more 'coordinated' in their vibrations than 'discoordinated'. Consequently it is natural to call the second term of (66) the *intra-cluster communicability* and the third term as the *inter-cluster* one (Estrada and Hatano, 2008). Mathematically, the difference between intra- and inter-cluster communicability is written as:

$$\Delta G_{pq}^{EA} = G_{pq}^{EA} - \phi_{1,A}(p)\phi_{1,A}(q)e^{\lambda_{1,A}}$$
$$= \sum_{j=2}^{\text{intracluster}} \phi_{j,A}(p)\phi_{j,A}(q)e^{\lambda_{j,A}} - \left| \sum_{j=2}^{\text{intercluster}} \phi_{j,A}(p)\phi_{j,A}(q)e^{\lambda_{j,A}} \right| \quad (67)$$

A community is then defined based on the communicability as a subset of nodes $C \subset V$ in the network $G = (V, E)$ for which the intracluster communicability is larger than the intercluster one for most of the nodes in $C$, which are then grouped according to a given quantitative criterion (Estrada and Hatano, 2008; 2009a; Estrada 2011). Several approaches have been proposed for identifying communities on the basis of the communicability function. For instance, the network can be transformed into a communicability graph, where two nodes are connected if and only if $\Delta G_{pq}^{EA} > 0$. Then, overlapped communities are identified as the cliques of this graph (Estrada and Hatano, 2008; 2009a). In FIG. 7a we illustrate the friendship network of a karate club studied by Zachary (1977) and its communicability graph (FIG. 6b). The method described before detected five communities (See FIG. 6c), three of which are highly overlapped ones. This may be one of the main drawbacks of this approach, which in general produces a large number of highly overlapped communities. This situation can be resolved in different ways, such as by considering hierarchical approaches based on the similarity between the communicability of pairs of nodes (Estrada, 2011) (see FIG 6d). However, when hierarchical methods as the ones proposed in (Estrada, 2011) are used the nice feature of having overlapped communities is lost.

a)                                              b)

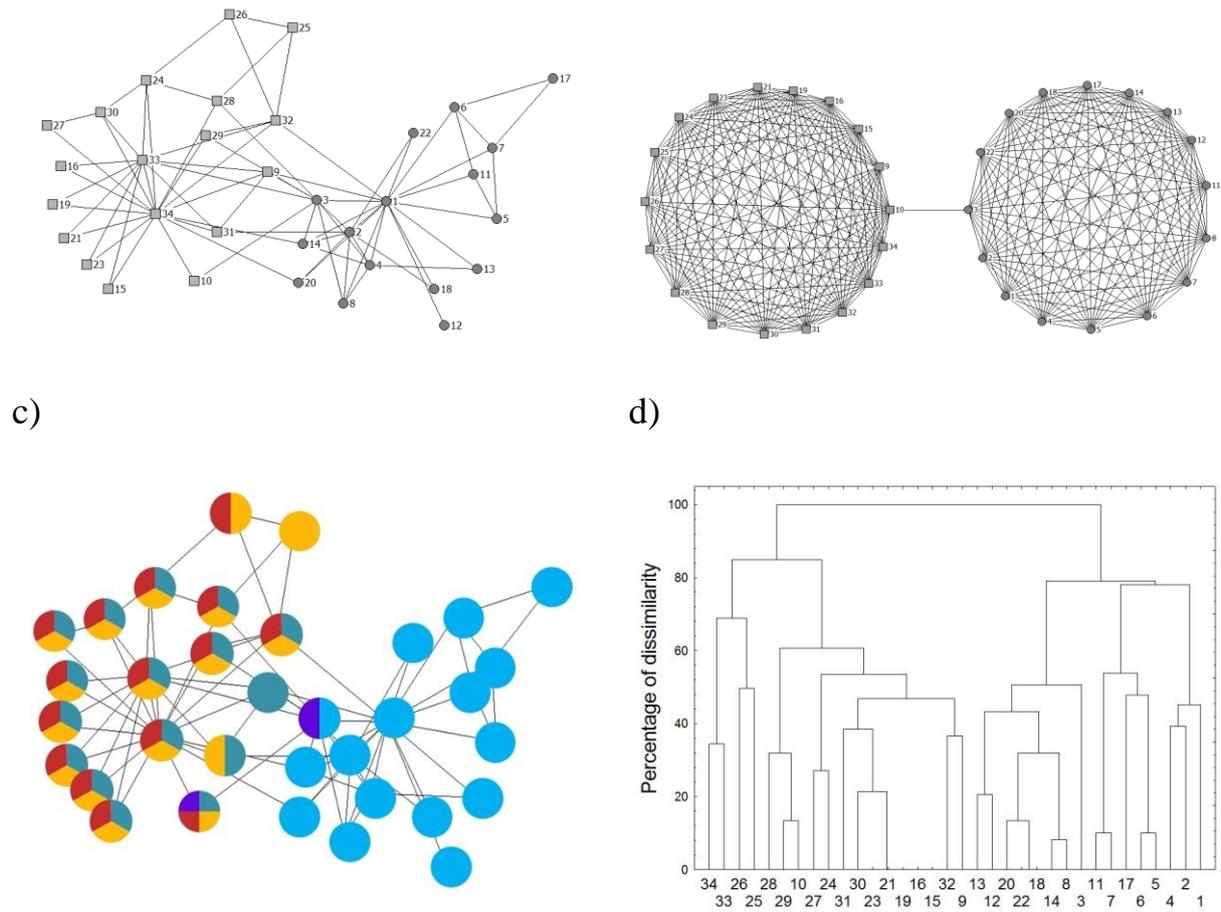

FIG. 6 (color online). Diagrammatic representation of the Zackary karate club network (a) and its communicability graph (b). c) Overlapped communities in the network represented in (a) and (d) the similarity among nodes used to detect hierarchical communities (see text).

The divorce between hierarchical and overlapping communities appears to be solved in a recent paper by Ma, Gao and Yong (2010), who present a new approach to community detection based on the communicability. The algorithm that they propose allows one to identify the overlapping and hierarchical community structure in complex networks more precisely than with other approaches presented in the literature, such as eigenvector-based methods or the nonnegative matrix factorization (NMF). The algorithm in (Ma *et al.*, 2010) uses several tunable factors, including the inverse temperature $\beta$ in the matrix exponential $e^{\beta A}$, an upper bound for the length of a short cycle, and the threshold for the density of cycles in a community. An appropriate choice of parameters seems to be crucial for obtaining good results, and it remains an open question how to select "good" parameters automatically.

Another important issue is how to exploit sparsity in the adjacency matrix so as to improve on the generic time complexity of $O(n^3)$ for a network with *n* nodes (see next section). Ma and Gao (2011) have compared several non-traditional spectral clustering methods for the detection of communities in complex networks. They have determined that the communicability-based approach achieves the best performance but is the slowest one (see further section on computability for an analysis).

Using their algorithm Ma *et al.* (2010) discovered several protein complexes in the yeast PPI. They identified many of these modules with functional categories included into MIPS (Mews *et al.*, 2002) and classified their complexes according to the number of functions the proteins in them share. For instance, in FIG. 7 we reproduce their results of modules, in which most of the proteins are involved in one, two and three functions.

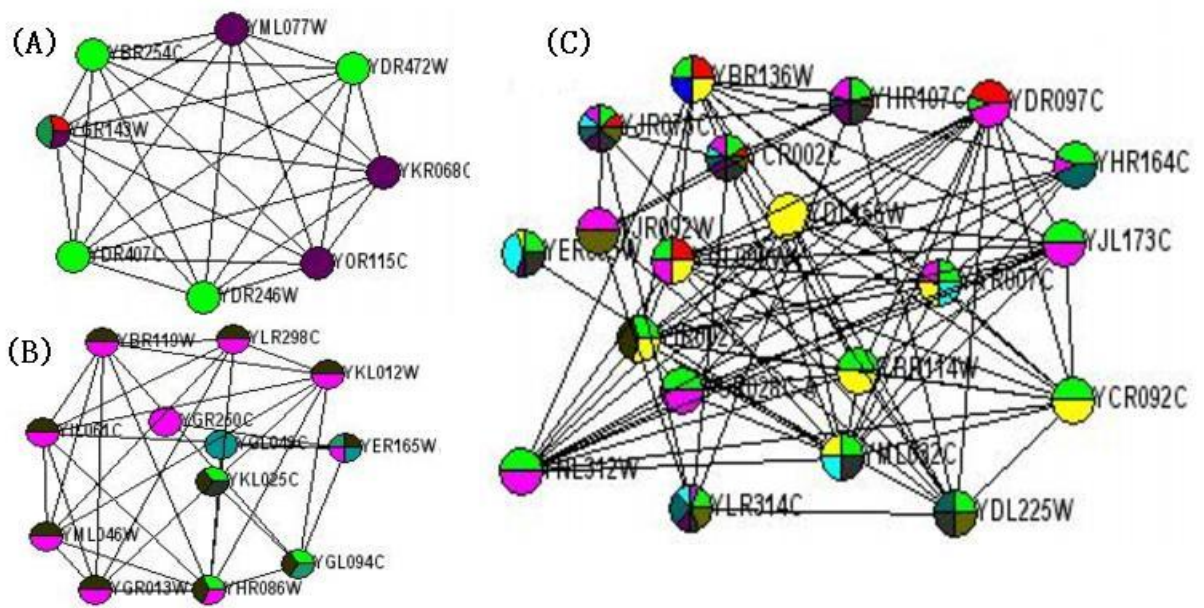

FIG. 7 (color online). Modules of the PPI network of yeast as found by the algorithm according to Ma *et al.* (2010). (A) Most proteins in these modules involve transportation; (B) proteins having two functions: transcription and protein binding; (C) a large module in which proteins have three functions: metabolism, DNA processing, and cell rescue. The figure is courtesy of Ma *et al.* (2010).

**C. Macroscopic analysis of networks**

Moving now on to the macroscopic analysis of complex networks, we find many examples of the use of the communicability, in particular for studying the robustness of complex networks in different contexts. The network robustness is a measure of the resilience of the network as a whole to the loss of nodes and links. The so-called '*natural connectivity*' has been used to study the robustness of several classes of artificial and real-world networks (Wu *et al.*, 2010a; b) using various strategies for removing links. The natural connectivity corresponds to the logarithm of the average Estrada index for a network:

$$\bar{\lambda} = \ln\left[EE(G)/n\right]. \tag{68}$$

It has been concluded that this index "*has strong discrimination in measuring the robustness of complex networks and exhibits the variation of robustness sensitively, even for disconnected networks*" (Wu *et al.*, 2010a). The interpretation of this index is straightforward in the context of statistical mechanics of the vibration of networks. For instance, it was pointed out that the entropy $S(G,\beta)$, the total energy $H(G,\beta)$ and Helmholtz free energy $F(G,\beta)$ of the network are given by (Estrada and Hatano, 2007)

$$S(G,\beta) = -k_B \sum_j \left[p_j\left(\beta\lambda_j - \ln EE\right)\right], \tag{69}$$

$$H(G,\beta) = -\sum_{j=1}^{n} \lambda_j p_j, \tag{70}$$

$$F(G,\beta) = -\beta^{-1} \ln EE, \tag{71}$$

where $p_j = e^{-\beta E_j}/EE$ is the probability that the network is found in a vibrational state of energy $E_j = -\lambda_{j,A}$. Here we have used $EE = EE(G,\beta)$. Then, the so-called '*natural connectivity*' of a network (Wu *et al.*, 2010a; b) can be rewritten for $\beta \equiv 1$ as

$$\bar{\lambda} = -\ln(n) - \left(-\ln\left[EE(G)\right]\right) = F(\bar{K}_n) - F(G), \tag{72}$$

where $\bar{K}_n$ stands for the complement of the complete graph, i.e., a graph with $n$ nodes and no link. This means that $\bar{\lambda}$ is the change of free energy of a hypothetical reaction in which all links of a given network are removed. We recall that $\Delta F = F_{\text{final}} - F_{\text{initial}}$. In other words it is the free energy gained by a network by having the connectivity pattern that it actually has.

A topic which can also be related to the network robustness is the identification of structural changes that produce significant disturbances in the functioning of the complex systems represented by networks. This is extremely important, for instance, in the science of complex materials where the challenge is to decipher their inherent structural design principles as they deform in response to external loads. Todesillas *et al.* (2010) have pioneered the study of dense granular materials represented as complex networks. Walker and Tordesillas (2010) have used the fact that when a force is applied to a material, the number of small cycles within the contact network decreases and longer cycles appear. They studied these transformations in granular materials represented by contact networks, using several network measures, including the subgraph centrality and a measure of bipartivity based on it. The spectral measure of bipartivity is the ratio of the self-communicability of a node based only on even closed walks to the total self-communicability of this node (Estrada and Rodríguez-Velazquez, 2005b):

$$b_S(G) = tr\left[\cosh(A)\right] / tr\left[\exp(A)\right] = \sum_{j=1}^{n} \cosh(\lambda_{j,A}) / EE(G). \tag{73}$$

Walker and Todesillas (2010) have found that the average subgraph centrality and the network bipartivity reflect the changes in the topology of the granular materials produced by the external strain (see FIG. 8a). The authors then studied a weighted version of the self-communicability by considering the magnitude of the normal force component between two particles as the weight of the corresponding link in the network. They determined that the behavior of the weighted subgraph centrality follows closely that of the shear stress and that the drops in this quantity coincide with the increase in the dissipation energy (Walker and Tordesillas, 2010). An example of the evolution of a buckling force chain on a small cluster as followed by the weighted subgraph centrality is shown in FIG. 8b. All in all, Walker and Tordesillas showed that the weighted subgraph centrality *"correlates strongly with nonaffine deformation and dissipation, spatially and temporally, and at both the mesoscopic and macroscopic level"*.

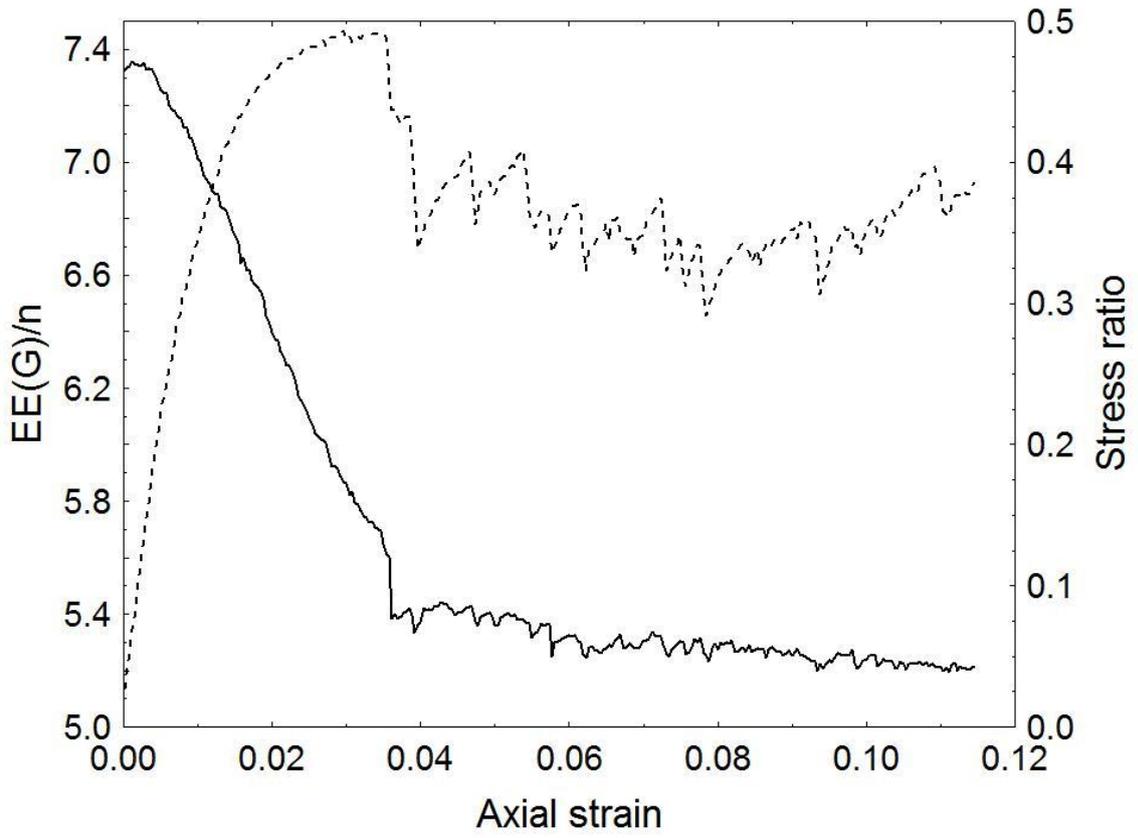

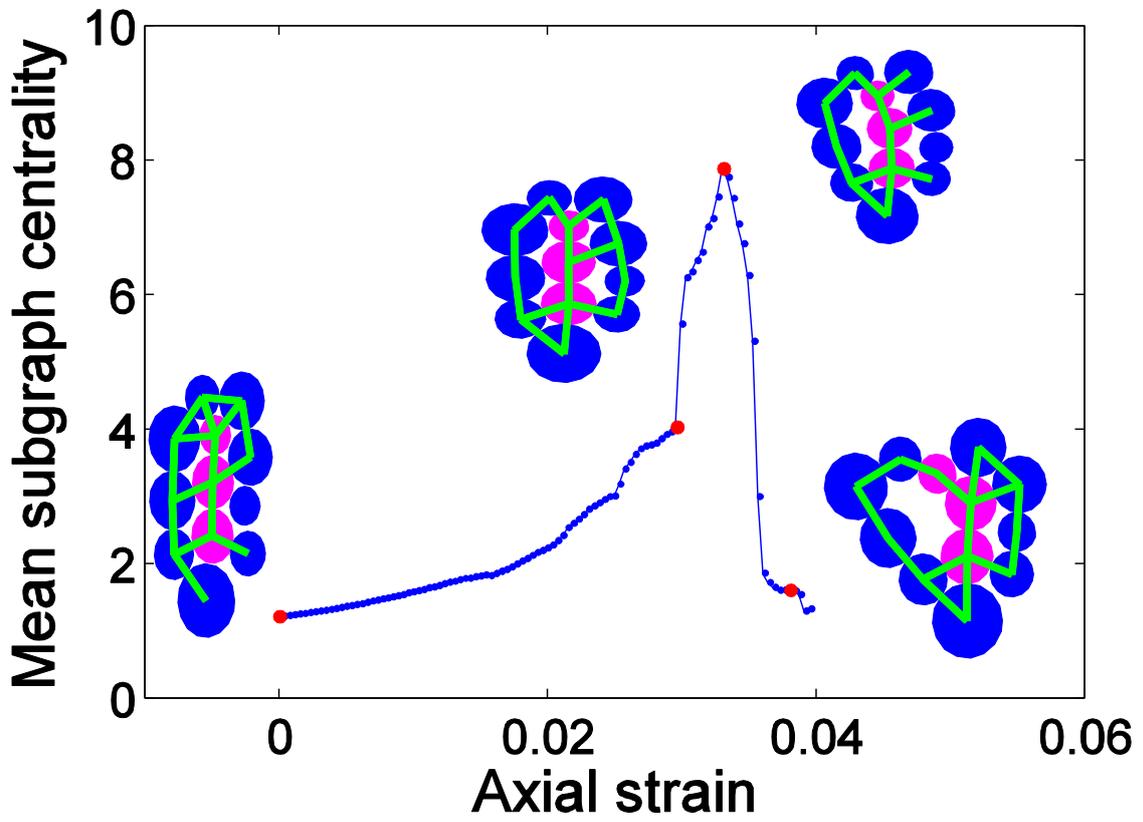

FIG. 8 (color online). (top panel) Average self-communicability (subgraph centrality) based on the exponential of the adjacency matrix for the global contact network for shear band particles normalized between 0 and 1. The shear stress is shown as a dotted line and the subgraph centrality as a solid one. (bottom panel) Evolution of a buckling force chain event. Configuration of the force chain and its confining neighbors, contact network and weighted subgraph centrality. The figure is courtesy of D. M. Walker and A. Tordesillas.

Another area that has attracted much attention in the study of complex networks is the study of anatomical and functional brain networks (Bullmore and Sporn, 2009; Johansen-Berg *et al.*, 2010; Sporn, 2011). A weighted communicability measure based on the normalized adjacency matrix was recently used by Crofts and Higham (2009) to study anatomical networks of human brains divided into 48 cortical and 8 subcortical regions. The networks were built from structural diffusion-weighted imaging data for 9 stroke patients at least six months following first, left hemisphere, subcortical stroke, and 10 age-matched control subjects. When considering data from the stroked hemisphere they discriminated stroke patients from controls in an effective way, which is "expected given the presence of a lesion and widespread degeneration in this hemisphere." However, in a further work, Crofts *et al.* (2011) studied 9 chronic stroke patients and 18 age-matched controls for whom brain networks were built by using diffusion MRI tractography. This time the communicability function was able to differentiate both groups by using information from the *contralesional* hemisphere, despite the absence of gross structural pathology in it. They found reduced communicability in brain regions surrounding the lesions in the affected hemisphere and around remote, but interconnected, homologue locations in the contralesional hemisphere (see FIG. 9).

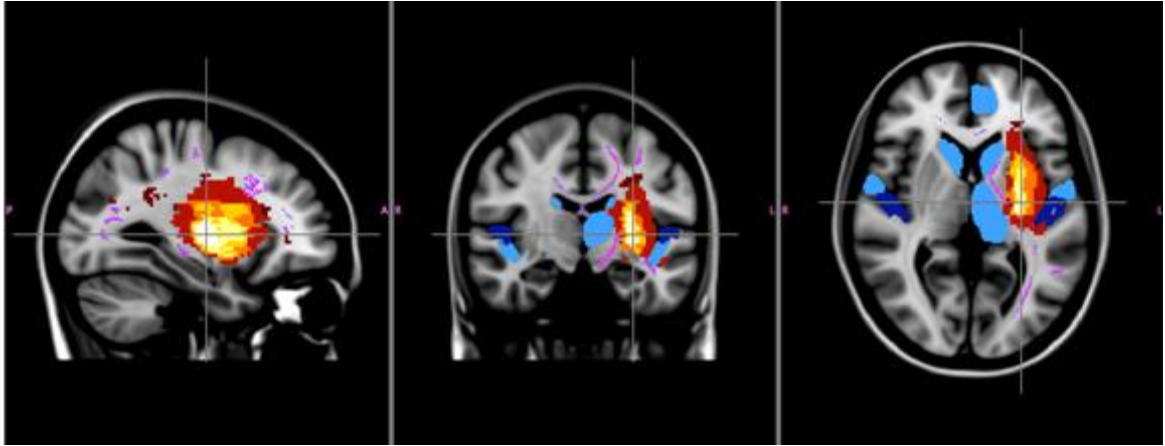

FIG. 9 (color online). Illustration of stroke lesions in human brains (red) and regions with reduced communicability. Reduced communicability is represented in blue and regions with increased communicability are represented in green. The figure courtesy of J. J. Crofts, reproduced with permission from Crofts *et al.* (2011).

**D. Multiscale analysis of networks**

A characteristic feature of complex systems is the difficulty in determining the borderlines of the system. The communicability function has been used to study the properties of networks known as good expansion, which allows determining whether a network is homogeneous enough as for 'expanding' its properties from a small sampling to the whole system. In order to explain this concept, let us consider a network in which we select an arbitrary subset of nodes containing no more than half the total number of nodes. Then, suppose that the number of links between nodes in the subset is approximately equal to the number of links between nodes in and out the subset. If this situation is repeated for any subset of the nodes in the network we say that the network is an expander or it has good expansion (GE) properties. This kind of networks does not contain structural bottlenecks; a bottleneck is a link or a node that after removal leaves the network disconnected. GE networks (GENs) have found many applications in a variety of fields (Hoory *et al.*, 2006).

A method of determining whether a network has GE properties has been designed on the basis of the self-communicability of a node (Estrada 2006c, d). The spectral scaling method uses the odd subgraph centrality $G_{pp}^{EA}(\text{odd})$, which is written in the following way:

$$G_{pp}^{EA}(\text{odd}) = \left[EC(p)\right]^2 \sinh(\lambda_1) + \sum_{j \geq 2}\left[\phi_{j,A}(p)\right]^2 \sinh(\lambda_j), \tag{74}$$

where $EC(p) = \phi_{1,A}(p)$ is the $p$ th component of the principal (Perron-Frobenius) eigenvector $\varphi_{1,A}$ corresponding to the largest eigenvalue $\lambda_{1,A}$ of the network, which is also known as the eigenvector centrality of the node $p$. Then, if a network has GE properties we can assume $\left[EC(p)\right]^2 \sinh(\lambda_1) \gg \sum_{j \geq 2}\left[\phi_{j,A}(p)\right]^2 \sinh(\lambda_j)$. That is, if the network has GE properties the translational movement of the network dominates over all vibrational states, which divides the network into many different communities as we have previously seen. In the case of a $k$-regular network it is known that a large spectral gap, i.e., the difference between the first and second largest eigenvalues of the adjacency matrix ($\lambda_1 - \lambda_2$), implies good expansion properties (Alon, 1986; Alon and Milman, 1985). In that case we can asume that the spectral gap is large enough so that: $\left[EC(p)\right]^2 \sinh(\lambda_1) \gg \sum_{j \geq 2}\left[\phi_{j,A}(p)\right]^2 \sinh(\lambda_j)$. Therefore, in the general case a GEN has odd-subgraph centrality that can be written as

$$G_{pp}^{EA}(\text{odd}) = \left[EC(p)\right]^2 \sinh(\lambda_1). \tag{75}$$

This means that the principal eigenvector of the network is directly related to the subgraph centrality in GENs according to the following spectral power-law scaling relationship:

$$EC(p) \propto A\left[G_{pp}^{EA}(\text{odd})\right]^{\eta}, \tag{76}$$

where $A = \left[\sinh(\lambda_1)\right]^{-0.5}$ and $\eta = 0.5$. This expression can be written in a log-log scale as

$$\log\left[EC(i)\right] = \log A + \eta \log\left[G_{pp}^{EA}(\text{odd})\right]. \tag{77}$$

Consequently, in a GEN a log-log plot of $EC(p)$ vs. $G_{pp}^{EA}(\text{odd})$ displays a perfect straight line fit with slope $\eta = 0.5$ and intercept $\log A$. Networks that do not possess GE properties will display large deviations from this perfect fit. This method has been used to classify complex networks into different universal structural classes (Estrada, 2007c) and has allowed the generation of algorithms for constructing networks that reproduce some of these structural classes (van Kerrenbroeck and Marinari, 2008). The problem of determining whether a network displays GE properties is of relevance in sampling networks. For instance, GENs are characterized by a large structural homogeneity across the scales of the network. Then, by sampling a relatively small part of the network we can make a good estimation of the general properties of the network as a whole.

Furthermore, the investigation of these properties is also important for generating realistic models of networks, for searching in networks as well as for the analysis of rumour spreading in networks. Recently, Malliaros and Megalooikonomou (2011) have studied several large social networks, three of them corresponding to collaboration networks and 6 online social networks with up to 1,134,890 nodes and 2,987,624 links. They have found that most of these social networks have GE properties according to the spectral scaling method but the two smallest networks display bad expansion properties (see FIG. 10). They have argued that these GE properties can be due to the large sizes of these systems, where it is difficult to find subsets of nodes that can be easily isolated. Another possibility is the fact that most of these networks are created over online social networking, which can facilitate the establishment of 'social' relationships between the agents (see for instance Dumbar, 1998). It is known for instance that online social networks differ in the connectivity patterns from those of more 'classical' collaboration networks (Hu and Wang, 2009).

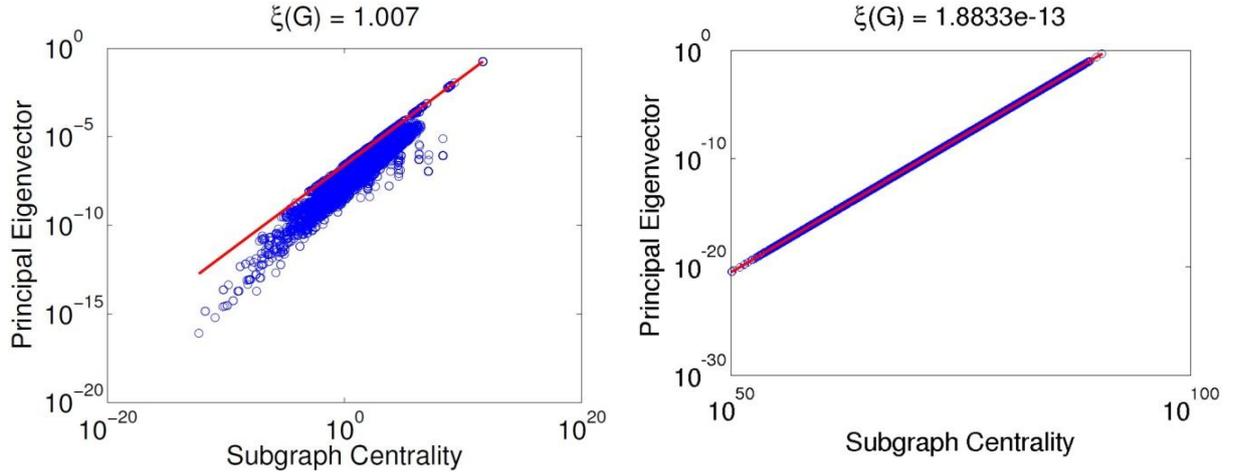

FIG. 10 (color online). Expansion properties of large social networks in which the principal eigenvector is plotted versus the self-communicability (subgraph centrality) in log-log scale and the values of the standard deviation $\xi(G)$ from perfect scaling are shown. The left plot corresponds to the collaboration network of co-authorship in the field of high-energy physics and the right one to the social network from Youtube site. The figure is courtesy of Mallarios and Megalooikonomou.

**E. Communicability at negative absolute temperature**

In all the previous examples the parameter $\beta$ has been assigned real positive values. That is, we have studied the behavior of the communicability functions at positive absolute temperatures. However, an interesting situation arises when we study the communicability function at negative absolute temperatures. The reader not accustomed to the concepts of thermal physics can find the use of negative absolute temperatures strange. Therefore, we provide here a short introduction to this concept based on the accounts of Ramsey (1956) and Baierlain (1999). The thermodynamic definition of the temperature for a system in equilibrium at a constant volume is given by:

$$\frac{1}{T} = \left(\frac{\partial S}{\partial U}\right)_{V,N}, \tag{78}$$

where $S$ and $U$ are the entropy and the internal energy, respectively, and $V$ and $N$ are the volume and the number of particles in the systems, which remains constant. If we consider a plot of $U$ versus $S$, the inverse temperature is defined as the slope of this curve at a given point. Consequently, if we consider a system of $n$ ideal paramagnets with spin $(1/2)\hbar$, there is a point of the minimum energy (a) which corresponds to the case where all spins are aligned with an external magnetic field. The point of the maximum energy (c) corresponds to an anti-alignment of all spins with respect to the external field. Both situations (a) and (c) have the minimal entropy as those systems are completely ordered, i.e., $U = 0$. However, there is an intermediate point (b) between (a) and (c) where one spin is up and its neighbors are down, which corresponds to the situation of the maximum entropy (see FIG. 11).

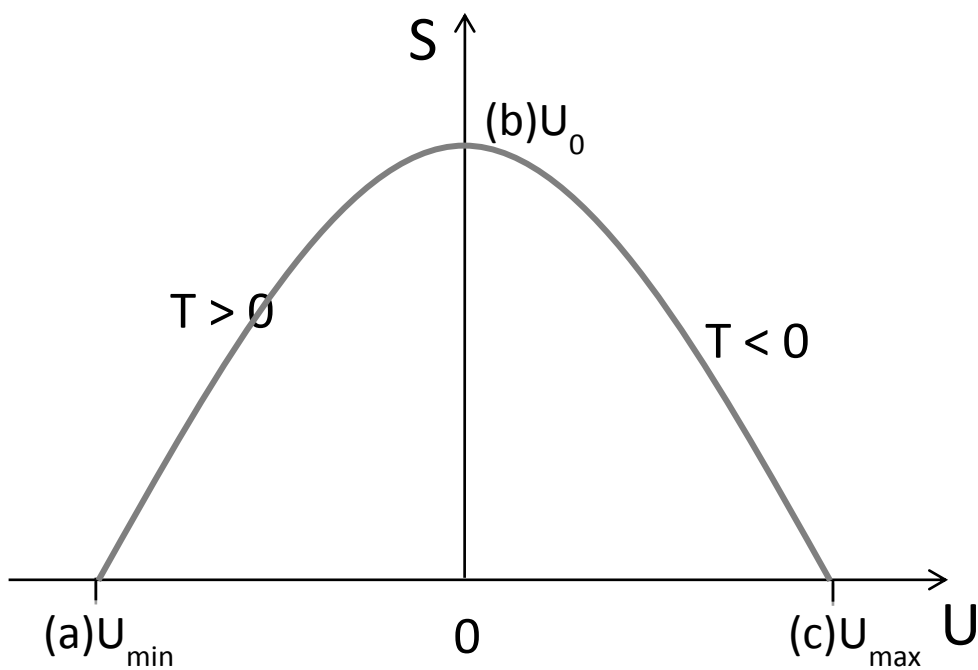

FIG. 11. Plot of the energy versus the entropy for a hypothetical system in equilibrium at constant volume.

Therefore, while the derivative at any point between (a) and (b) is positive, indicating that the temperature has positive values, the slope of the curve between (b) and (c) is negative and so is the temperature. Then, the absolute temperature runs from cold to hot as: $0K, \cdots, +300K, \cdots, +\infty K, \cdots, -\infty K, \cdots, -300K, \cdots, -0K$, which means that absolute negative temperatures are hotter than positive ones.

In order to study the communicability function at negative absolute temperatures, let us start by writing it in the following form:

$$G_{pq}^{EA}(\beta) = \sum_{\lambda_j < 0} \phi_j(p)\phi_j(q) e^{\beta\lambda_j} + \sum_{\lambda_j = 0} \phi_j(p)\phi_j(q) e^{\beta\lambda_j} + \sum_{\lambda_j > 0} \phi_j(p)\phi_j(q) e^{\beta\lambda_j}. \qquad (79)$$

While the eigenvectors associated with positive eigenvalues make contributions to the partition of the network into communities or quasi-cliques, the eigenvectors associated with negative ones make contributions to the partition of the network into quasi-bipartite clusters. Then, for $\beta < 0$, the first term of (79) makes the largest contribution to the communicability, such that (Estrada *et al.*, 2008):

$$G_{pq}^{EA}(\beta < 0) \approx \sum_{\lambda_j < 0}^{n} \varphi_j(p)\varphi_j(q) e^{-|\beta|\lambda_j}, \qquad (80)$$

which means that for $G_{pq}^{EA}(\beta < 0)$ the network is partitioned into quasi-bipartite clusters. This can be easily seen by considering

$$e^{-|\beta|A} = I - |\beta|A + \frac{(|\beta|A)^2}{2!} - \frac{(|\beta|A)^3}{3!} + \cdots, \qquad (81)$$

which can be expressed in terms of the hyperbolic functions as

$$e^{-|\beta|A} = \cosh(|\beta|A) - \sinh(|\beta|A). \qquad (82)$$

As we have previously seen, the term $\left[\cosh(|\beta|A)\right]_{pq}$ represents the weighted sum of the number of walks of even length connecting nodes $p$ and $q$ in the network. Similarly, $\left[\sinh(|\beta|A)\right]_{pq}$ represents the weighted sum of the number of walks of odd length connecting nodes $p$ and $q$. Then, if we consider a bipartite graph in which $p$ and $q$ are nodes which are in two different partitions of the network, it is straightforward to realize that there are no walks of even length starting and $p$ at ending at $q$ in the graph. Consequently,

$$G_{pq}^{EA}(\beta<0) = \left[-\sinh(|\beta|A)\right]_{pq} < 0. \tag{83}$$

On the other hand, if $p$ and $q$ are in the same partition of a bipartite network we can see that there is no walk of odd length connecting them due to the lack of odd cycles in the bipartite graph, which makes

$$G_{pq}^{EA}(\beta<0) = \left[\cosh(|\beta|A)\right]_{pq} > 0. \tag{84}$$

Thus, it is possible to adapt the methods and algorithms previously described to identify communities in networks in order to identify network bipartitions. These methods have been used for undirected networks (Estrada *et al.*, 2008) in which bipartitions have been detected for a variety of real-world systems. For instance, in FIG. 12 we illustrate the two main partitions detected by using the communicability function at a negative absolute temperature for the PPI network of the archae bacterium *A. fulgidus*. The study of bipartitions in complex directed networks was also accomplished by using a modification of the communicability function (Crofts *et al.*, 2010).

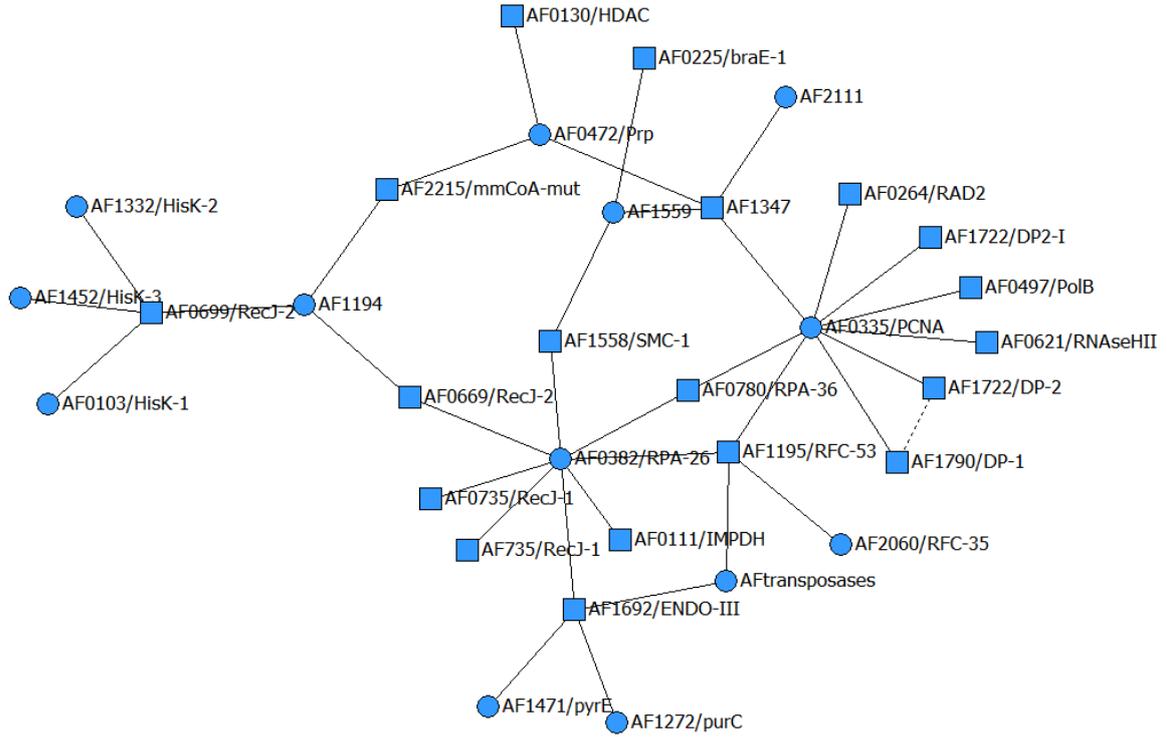

FIG. 12 (color online). Bipartition of the PPI network of *A. fulgidus* obtained by means of the communicability function at a negative absolute temperature. The nodes in one partition are represented by squares and those in the other as circles. Inter-partition links are represented by solid lines and the only one intra-partition link is represented as a discontinuous line.

## VI. COMMUNICABILITY AND LOCALIZATION IN COMPLEX NETWORKS

### A. Generalities

Locality is an important property of a large class of physical systems. For example, in quantum chemistry and solid-state physics, the locality (also known as nearsightedness, see (Des Cloizeaux, 1964; Kohn, 1996; Le Bris, 2005)) can be interpreted as the lack of long-range correlations between the components of the system, which can be modeled by sites connected by bonds in a more or less regular lattice. This means that with high probability, a small perturbation at one site of the lattice will only be felt locally (i.e., by sites in a small neighborhood).

Mathematically, this property manifests itself as fast off-diagonal decay in the density matrix describing the system at hand; see, e.g., (Benzi *et al.*, 2010). The locality is present in insulators as well as in metallic systems at sufficiently high electronic temperatures. One important consequence of the locality is that it enables the development of $O(n)$ algorithms for electronic structure computations, i.e., algorithms the asymptotic complexity of which scales linearly in the size of the system (Goedeker, 1999). Note that traditional algorithms based on diagonalization of the Hamiltonian, in contrast, scale like $n^3$.

The locality is also of great importance in the area of quantum information theory, where it has been used, for example, to establish *area laws* for the entanglement entropy; see, e.g., (Cramer and Eisert, 2006; Cramer *et al.*, 2006; Hastings, 2004; Hastings and Koma, 2006; Schuch, 2007; Schuch *et al.*, 2006) and especially (Eisert *et al.*, 2010) for a comprehensive survey.

Conversely, the *absence* of locality can be thought of as the presence of long-range correlations throughout the system. In such a system, a small local perturbation will be felt globally. In quantum chemistry this happens for conductors, e.g., metallic systems at zero or very low electronic temperatures. In these systems, the entries of the density matrix decay very slowly.

It should be obvious that the locality (or the lack of it) is also very important in the study of complex networks. In this section we apply general results on exponential decay in matrix functions (Benzi *et al.*, 2010; Benzi and Golub, 2009; Benzi and Razouk, 2007) to the study of a type of locality in complex networks. Specifically, we use the communicability as a way to measure the correlations between nodes in a network. Thus, fast decay in $e^{\beta H}$ (see below for details) will be interpreted as a sign of localization, meaning the absence of strong long-range correlations among the nodes; conversely, slow decay in $e^{\beta H}$ (or the lack of decay) will be interpreted as a sign of a strongly connected network, in which even small disturbances propagate easily to the entire network. Here $H$ may denote either the adjacency matrix or the graph Laplacian of the network.

Just as in the case of electronic structure computations, the locality is not only conceptually important but may also lead in some cases to greatly reduced computational effort, for instance in computing communicabilities or other network properties expressible as matrix functions. It is, however, important to realize that such decay properties will be present only in some networks, but not in others. For instance, in a small-world network we cannot expect most communicabilities to be negligibly small, generally speaking. On the other hand, other types of networks, such as regular lattices or highway networks can be expected to exhibit strong locality. The actual rate of decay is affected by properties such as the maximum degree of a node in the network and the inverse temperature $\beta$.

Let us consider for instance the 1997 version of the Internet at Autonomous System (AS) formed by 3015 nodes and 5156 links. Despite this network is sparse, the maximum communicability $G_{pq}^{EA}$ between a pair of nodes is $\sim 10^{13}$. The minimum communicability is, however, only 3.1. Then, by normalizing the communicability matrix it is easy to realize that the minimum communicability is negligibly close to zero, i.e., $\sim 10^{-13}$ and can be excluded from the calculations. The average normalized communicability in this version of the AS Internet is $4.07 \times 10^{-4}$; 16.7% of the pairs of nodes have a communicability smaller than $10^{-6}$.

There are networks where the number of negligible entries is more significant than this; in the network of 616 injecting drug users (IDUs) in Colorado Spring, $G_{pq}^{EA} < 10^{-6}$ for almost 30% of the pairs of nodes. We remark that there are some networks in which this situation is not found at all; in the network of 1586 corporate directors of the top 500 US corporations, only 0.4% of the pairs of nodes have communicability below $10^{-6}$. These cases of 'locality' (or lack thereof) of the communicability in a group of nodes in the network are telling us something about the structure of these networks. For instance, in the case of the IDU network there is a central core dominating most of the communicability of the network as can be seen in FIG. 13. These individuals are central in the communication with the rest of the network and could be important targets of educational or health campaigns.

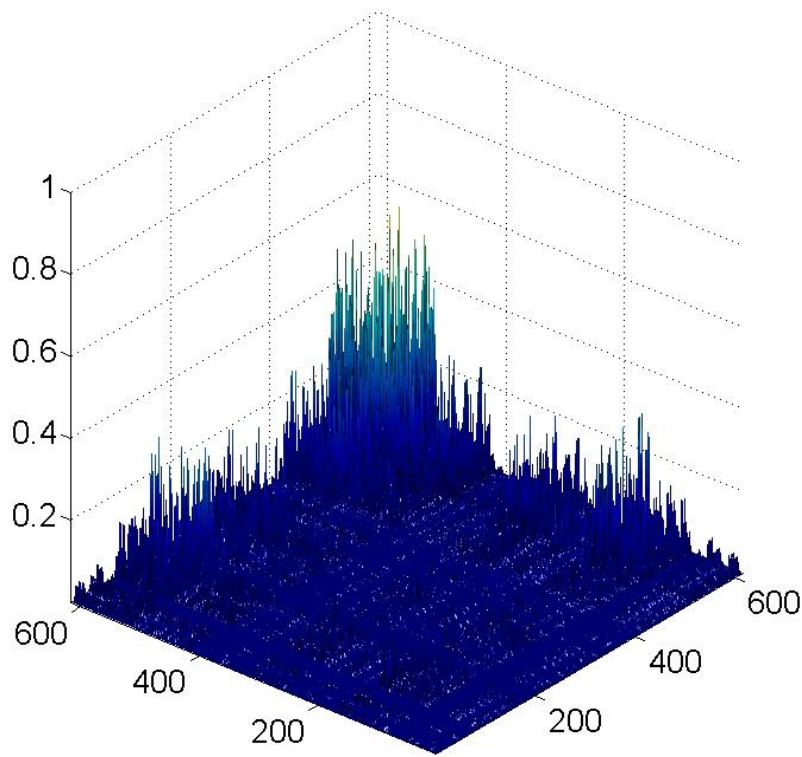

FIG. 13 (color online). Illustration of the normalized communicability between pairs of individuals in a network of IDUs in Colorado Spring. Individuals are identified by numbers in the $x$ and $y$ axes and their communicability is given as the values of the $z$ axis.

## B. Exponential decay in communicability

In the case of a symmetric matrix $M$ with the spectrum contained in $I = [-1, 1]$ and for a matrix function $f$ which is analytic on a region containing $I$, it has been proved that the off-diagonal entries of this matrix function are bounded as

$$\left| f(M)_{pq} \right| \leq C e^{-\tau d_{pq}} \text{ for all } p \neq q, \tag{85}$$

where $C > 0$, $\tau > 0$ and $d_{pq}$ denotes the shortest-path distance between the nodes $p$ and $q$ in the undirected and unweighted graph associated with $M$. This result reveals the same information as we have analyzed in section 2 of this work (see Eq. (20) for instance) for the special case of the path graph. Basically, as the shortest-path separation between two nodes in a network increases, their communicability vanishes. Note, however, that the actual decay is not monotonic in general.

Now, let us consider the adjacency matrix $A$ of an undirected network and let $k_{\max}$ be the maximum degree of the network. Then, we can normalize the adjacency matrix as we have done previously producing $\tilde{A} = k_{\max}^{-1} A$ which has spectrum contained in the interval $I = [-1, 1]$. Then, we can obtain decay bounds for the exponential of $A$ by computing bounds for the matrix function $f(\tilde{A}) = e^{k_{\max} \tilde{A}}$. The bounds are of the form

$$G_{pq}^{EA} \leq C e^{-\tau d_{pq}} \text{ for all } p \neq q, \tag{86}$$

where now $C$ and $\tau$ depend only on $k_{\max}$. We will show later that a larger value of $k_{\max}$ results in a larger constant $C$, and therefore a slower decay.

A shift and scaling is needed for the Laplacian to have spectrum in the interval $I = [-1, 1]$. We start by denoting

$$\hat{L} = \left(\frac{2}{\lambda_{\max,L}}\right) L - I \tag{87}$$

the shifted and scaled Laplacian. Observing that

$$L = \left(\frac{\lambda_{\max,L}}{2}\right) \hat{L} + \left(\frac{\lambda_{\max,L}}{2}\right) I, \tag{88}$$

we immediately have

$$e^{-L} = e^{-\lambda_{\max,L}/2} e^{-\lambda_{\max,\hat{L}}/2}. \tag{89}$$

Thus, in order to obtain bounds on $G_{pq}^{EL}$ we compute bounds for the matrix function $f(\tilde{A}) = e^{\frac{-\lambda_{\max,L}}{2}\tilde{A}}$, and then multiply the bounds by the constant factor $e^{\frac{-\lambda_{\max,L}}{2}}$.

Consider now the particular matrix function $f(\tilde{A}) = e^{t\tilde{A}}$, where $t > 0$ and $\tilde{A}$ has spectrum in $I = [-1, 1]$. Applying the bounds (85) and (86) to this function leads to the upper bound

$$\left(e^{t\tilde{A}}\right)_{pq} \leq C(t) e^{-\tau d_{pq}}, \ p \neq q, \tag{90}$$

where (Benzi and Golub, 1999)

$$C(t) = \frac{2\chi e^{t\kappa_1}}{\chi - 1}, \ \tau = 2\ln\chi. \tag{91}$$

Here $\chi = \kappa_1 + \kappa_2 = \kappa_1 + \sqrt{\kappa_1^2 - 1}$, where $\kappa_1 > 1$ and $\kappa_2 > 0$ are the semi-axes of an ellipse $\Omega$ with foci at the points $-1$ and $+1$. The matrix function $f$ is analytic on the interior of $\Omega$ and continuous on it. The bound (90) can be explicitly evaluated for any given value of $\kappa_1 > 1$. Note that if $t < 0$, the factor $e^{t\kappa_1}$ in (91) is replaced by $e^{-t\kappa_1}$. Also note that owing to the presence of the factor $e^{t\kappa_1}$ in (91), the bounds for the entries of $e^A$ are larger for a larger value of $k_{max}$ (just take $t = k_{max}$).

If we consider the inverse temperature $\beta$ we can obtain the following bound for the Laplacian-based communicability function

$$G_{pq}^{EL} \leq C(\beta) e^{-\tau d_{pq}}, \quad p \neq q, \tag{92}$$

where

$$C(\beta) = \frac{2\chi}{\chi - 1} e^{\beta \lambda_{max,L}(\kappa_1 - 1)/2}, \quad \tau = 2\ln \chi. \tag{93}$$

Note that $C(\beta)$ increases to infinity as $T \to 0$, and decreases to 1 as $T \to \infty$. In the zero-temperature limit the bound deteriorates and no decay is observed, which is consistent with the observations above. In the limit $T \to \infty$ the right-hand side in the bound (90) tends to

$$\frac{2\chi}{\chi - 1} e^{-\tau d_{pq}}, \tag{94}$$

where $\tau = 2\ln \chi$ and $\chi > 1$ is arbitrary. Taking $\chi$ sufficiently large, the above quantity can be made smaller than any prescribed $\varepsilon > 0$, showing that in the limit $T \to \infty$ the off-diagonal entries of $e^{-\beta L}$ are all zero: therefore, our bounds capture the correct limiting behavior for both $T \to 0$ and $T \to \infty$.

In a similar way we can obtain the bound for the adjacency-based communicability,

$$G_{pq}^{EA} \leq C_A(\beta) e^{-\tau d_{pq}}, \quad p \neq q \tag{95}$$

with

$$C_A(\beta) = \frac{2\chi}{\chi - 1} e^{k_{\max} \kappa_1 \beta}, \quad \tau = 2\ln \chi. \tag{96}$$

As one would expect, the bound increases upon increasing $k_{\max}$ and increasing $\beta$ (or decreasing the temperature). In the limit $T \to \infty$ we again find $G_{pq}^{EA} \to 0$.

## VII. COMPUTABILITY OF COMMUNICABILITY FUNCTIONS

### A. Analytical results

Communicability functions have been incorporated into some computational tools for the analysis of complex networks. For instance, a Hub Objects Analyzer (Hubba) (Lin *et al.*, 2008) designed as a web-based service for exploring important nodes in an interactome network and BrainNetVis (Christodoulou *et al.*, 2011), a tool for both quantitative and qualitative network measures of brain interconnectivity, incorporate the subgraph centrality as a standard measure for the analysis of nodes in networks. A more general computational toolbox, CONTEST (Taylor and Higham, 2009), contains a series of Matlab utilities for generating and analyzing various types of networks and incorporates several communicability-based functions for the analysis of complex networks.

When implementing communicability functions for large complex networks a fundamental question that arises is the efficiency of the algorithm selected for computing the matrix functions. As a consequence it has been stated here and there that the subgraph centrality and the communicability are difficult to compute for large networks. There have been approaches as the ones described in the previous section based on the truncation of the eigenvalues of the adjacency matrix for which we do not know the error of the approximation. We present in this section a critical review of these computational approaches to give the reader a better understanding of what to do and what not to do when computing communicability functions in large complex networks.

Several approaches are available for computing the matrix exponential. Use of the power series expansion (8) to find approximations to $e^A$ in general cannot be recommended; see (Moler and van Loan, 1978; 2003). A frequently used approach is based on the eigendecomposition (9); see, e.g., (Dronen and Lv, 2011). This approach requires $O(n^2)$ space and $O(n^3)$ arithmetic operations for a graph with $n$ nodes; furthermore, the sparsity in $A$ is not exploited in this approach.

One of the most efficient and accurate available methods is the one based on the Padé approximation combined with the scaling and squaring method (Higham, 2005; 2008). This method, implemented in Matlab by the expm function, is nowadays the most widely used one. Its complexity is also $O(n^2)$ storage and $O(n^3)$ arithmetic operations, and the sparsity in **A** does not appear to be exploited in available implementations. (The sparsity in $A$ is exploited in codes that compute the action of the matrix exponential on a vector: $v = e^A b$; however, this problem is somewhat different from the one that we are interested in here.)

It is important to note that in many applications, it is not required to compute *all* entries in the matrix exponential (or in other functions of $A$ or $L$). In particular, for computing the graph centralities of the nodes in a graph (or the Estrada index) only the main diagonal of $e^A$ is required. However, neither the diagonalization (eigendecomposition) approach, nor the scaling and squaring method are able to take advantage of this to reduce computational costs. Moreover, for very large networks it is generally not feasible to compute *all* the communicabilities; instead, one may be interested in computing only the average communicability of each node in the network, or of a subset of nodes. Again, this can be easily done without computing all the entries in $e^A$, as we show below.

Efficient and accurate methods of bounding and estimating arbitrary entries in a matrix function $f(A)$ have been developed by Golub, Meurant and collaborators (see Golub and Meurant, 2010 and references therein) and were first applied to problems of network analysis by Benzi and Boito (2010) (see also Bonchi *et al.*, 2011). Here we give a brief description of these methods, referring the reader to (Benzi and Boito (2010)) for further details. Consider the eigendecomposition $A = Q\Lambda Q^T$ and $f(A) = Qf(\Lambda)Q^T$, where $Q = [\varphi_1, \ldots, \varphi_n]$. Given the vectors $u$ and $v$, we have

$$u^T f(A)v = u^T Q f(\Lambda) Q^T v = w^T f(\Lambda) z = \sum_{i=1}^{n} f(\lambda_i) w(i) z(i), \tag{97}$$

where $w = Q^T u$ and $z = Q^T v$. In particular, for $f(A) = e^A$ we obtain

$$u^T e^A v = \sum_{i=1}^{n} e_i^\lambda w(i) z(i). \tag{98}$$

Choosing $u = v = e_p$ (the vector with the $p$ th entry equal to 1 and all the remaining ones equal to 0) we recover the well-known expression for the subgraph centrality of node $p$:

$$EE(p) = e_p^T e^{\mathbf{A}} e_p = \sum_{i=1}^{n} e_i^\lambda \left[\phi_i(p)\right]^2. \tag{99}$$

Likewise, choosing $u = e_p$ and $v = e_q$ we obtain the usual expression for the communicability between node $p$ and node $q$:

$$G_{pq} = e_p^T e^{\mathbf{A}} e_q = \sum_{j=1}^{n} \phi_j(p) \phi_j(q) e^{\lambda_j}. \tag{100}$$

Let now 1 denote the column vector with all entries equal to 1. The average communicability for the node $p$ can be computed as

$$\langle C(p) \rangle = \frac{1}{n-1} \left[ 1^T e^A e_p - e_p^T e^A e_p \right]. \tag{101}$$

This shows that this quantity can be evaluated once the two bilinear forms $1^T e^A e_p$ and $e_p^T e^A e_p$ have been computed.

Hence, the problem is reduced to evaluating bilinear expressions of the form $u^T f(A) v$. The key insight is that such bilinear forms can be thought of as Riemann-Stieltjes integrals with respect to a (signed) spectral measure:

$$u^T f(A) v = \int_a^b f(\lambda) d\mu(\lambda), \qquad \mu(\lambda) = \begin{cases} 0, & \text{if } \lambda < a = \lambda_n, \\ \sum_{j=1}^{i} w(j) z(j), & \text{if } \lambda_{i+1} \leq \lambda < \lambda_i, \\ \sum_{j=1}^{n} w(j) z(j), & \text{if } b = \lambda_1 \leq \lambda. \end{cases} \tag{102}$$

This integral can be approximated by means of a Gauss-type quadrature rule:

$$\int_a^b f(\lambda) d\mu(\lambda) = \sum_{j=1}^{r} c_j f(t_j) + \sum_{k=1}^{s} v_k f(\tau_k) + R[f], \tag{103}$$

where the nodes $\{t_j\}_{j=1}^r$ and the weights $\{c_j\}_{j=1}^r$ are unknown, whereas the nodes $\{\tau_k\}_{k=1}^s$ are prescribed. We have

- $s = 0$ for the Gauss rule,

- $s = 1$, $\tau_1 = a$ or $\tau_1 = b$ for the Gauss-Radau rule,

- $s = 2$, $\tau_1 = a$ and $\tau_2 = b$ for the Gauss-Lobatto rule.

For certain matrix functions, including the exponential and the resolvent, these quadrature rules can be used to obtain lower and upper bounds on the quantities of interest;

adding quadrature nodes leads to tighter and tighter bounds, which converge to the true values. The evaluation of these quadrature rules is reduced to the computation of orthogonal polynomials via a three-term recurrence, or, equivalently, to the computation of entries and spectral information on a certain tridiagonal (Jacobi) matrix via the Lanczos algorithm. Here we briefly recall how this can be done for the case of the Gauss quadrature rule, when we wish to estimate the $i$ th diagonal entry of $f(A)$. It follows from (103) that the quantity of interest has the form $\sum_{j=1}^{r} c_j f(t_j)$. The nodes and weights can be efficiently computed using the Golub and Welsch QR algorithm, see (Golub and Meurant, 2010). Alternatively, one can use the following relation (Theorem 3.4 in (Golub and Meurant, 2010)):

$$\sum_{j=1}^{r} c_j f(t_j) = e_1^T f(J_r) e_1, \tag{104}$$

where

$$J_r = \begin{pmatrix} \omega_1 & \gamma_1 & & & \\ \gamma_1 & \omega_2 & \gamma_2 & & \\ & \ddots & \ddots & \ddots & \\ & & \gamma_{r-2} & \omega_{r-1} & \gamma_{r-1} \\ & & & \gamma_{r-1} & \omega_r \end{pmatrix} \tag{105}$$

is a tridiagonal matrix whose eigenvalues are the Gauss nodes, whereas the Gauss weights are given by the squares of the first entries of the normalized eigenvectors of $J_r$. The entries of $J_r$ are computed using the Lanczos algorithm. The initial vectors are $x_{-1} = 0$ and $x_0 = e_i$. The iteration goes as follows:

$$\begin{aligned} \gamma_j x_j &= r_j = (A - \omega_j I) x_{j-1} - \gamma_{j-1} x_{j-2}, \quad j = 1, 2, \cdots \\ \omega_j &= x_{j-1}^T A x_{j-1}, \\ \gamma_j &= \|r_j\| = r_j^T r_j^{1/2}. \end{aligned} \tag{106}$$

In practice, a slightly different implementation due to Paige is preferred for numerical reasons, see (Golub and Meurant, 2010).

For small $r$, i.e., for a small number of Lanczos steps, computing the (1,1) entry of $f(J_r)$ is inexpensive. The main cost in estimating one entry of $f(A)$ with this approach is associated with the sparse matrix-vector multiplies in the Lanczos algorithm applied to the adjacency matrix $A$. If only a small, fixed number of iterations is performed for each diagonal element of $f(A)$, as is usually the case, the computational cost (per node) is at most $O(n)$ for a sparse graph, resulting in a total cost of $O(n^2)$ for computing the subgraph centrality of every node in the network. This theoretical worst case applies for instance to small-world networks; for networks the diameter of which is not small, a careful sparse-matrix-sparse-vector implementation leads to an $O(\bar{k})$ complexity per node, where $\bar{k}$ is the average degree of a node in the network. This translates to an overall cost of $O(n)$ for computing all the subgraph centralities for a sparse network. The prefactor in the $O(n)$ and $O(n^2)$ estimates may be large, meaning that the quadrature rule-based approach will be faster than traditional $O(n^3)$ methods only for sufficiently large $n$. The break-even point will depend on the particular type of network being considered, but in our experience it can occur for $n$ as small as a few hundreds. Of course, the larger $n$, the greater the savings realized with the quadrature rule-based approach.

Furthermore, since each subgraph centrality can be computed independently of the others, parallelization of the computation will result in a drastic reduction of computing times, thus enabling the analysis of huge networks on massively parallel systems, at least in principle. We are not aware of any other approach to computing subgraph centralities with similar characteristics.

The case $u \neq v$ can be handled either by the nonsymmetric Lanczos process (Golub and Meurant, 2010), or by means of the following *polarization identity*:

$$u^T f(A) v = \frac{1}{2}\left[(u+v)^T f(A)(u+v) - u^T f(A) u - v^T f(A) v\right]. \tag{107}$$

For the case $f(A) = e^A$, $u = e_p$ and $v = e_q$ we obtain the following expression for the communicability between nodes $p$ and $q$:

$$G_{pq} = \frac{1}{2}\left[(e_i + e_j)^T e^{\mathbf{A}} (e_i + e_j) - EE(p) - EE(q)\right], \tag{108}$$

showing that once the subgraph centralities have been computed, only one additional quadratic form must be evaluated in order to compute $G_{pq}$. Hence, the cost of computing the communicability between a pair of nodes is of the same order as that of computing the subgraph centrality of a node.

For a very large network, computing the communicabilities between *all* pair of nodes would likely be too expensive. In this case computing the average communicabilities may be sufficient, depending on the problem; in other cases it may be sufficient to compute the communicabilities for a subset of nodes, or only those communicabilities that are not *a priori* known to be below a certain threshold (see discussion on decay).

As already mentioned, computing the average communicability for node $p$ requires evaluating the bilinear form $1^T e^A e_p$. This can be computed with the nonsymmetric Lanczos process, or alternatively with the symmetric Lanczos process via the alternative polarization identity

$$1^T e^A e_p = \frac{1}{4}\left[(1 + e_p)^T e^{\mathbf{A}} (1 + e_p) - (1 - e_p)^T e^{\mathbf{A}} (1 - e_p)\right]. \tag{109}$$

Hence, the cost of computing the average communicability of a node is of the same order as computing the subgraph centrality of a node. We note that all the average communicabilities can also be computed in parallel.

**B. Numerical experiments**

Here we present the results of computations on a set of small-world networks generated using the CONTEST toolbox (Taylor and Higham, 2009). The networks are obtained from an underlying regular lattice consisting of a ring, in which each node is connected to four neighbors (two on each side). A shortcut to a randomly chosen node in the network is added to each node in turn with probability $p = 0.1$, with self-links and repeated links removed at the end of the process.

First we consider a set of networks of $n$ nodes with $1000 \leq n \leq 4000$, with increments of 200. We compute the subgraph centralities of all nodes in the networks and then sum them to obtain the Estrada index. The results are shown in FIG. 14. As expected, the computational times with the eigendecomposition and with the Matlab function *expm* scale roughly like $n^3$. Even without a sophisticated implementation, our results show that the quadrature rule-based approach is systematically faster for networks of size greater than $n \approx 2000$. In this graph the time for the quadrature rule-based calculations appear to scale roughly linearly with $n$, rather than with the expected (quadratic) complexity. This is probably due to the fact that for such relatively small sparse problems, the computational time is dominated by non-floating point operations (including indexing, memory references, etc.), which scale roughly linearly for the range of problems of sizes considered here. This is confirmed by the plot in inset of FIG. 14, where we present results for $n = 1000, 5000, 10000$. This plot indicates a quadratic growth in computing times for sufficiently large graphs.

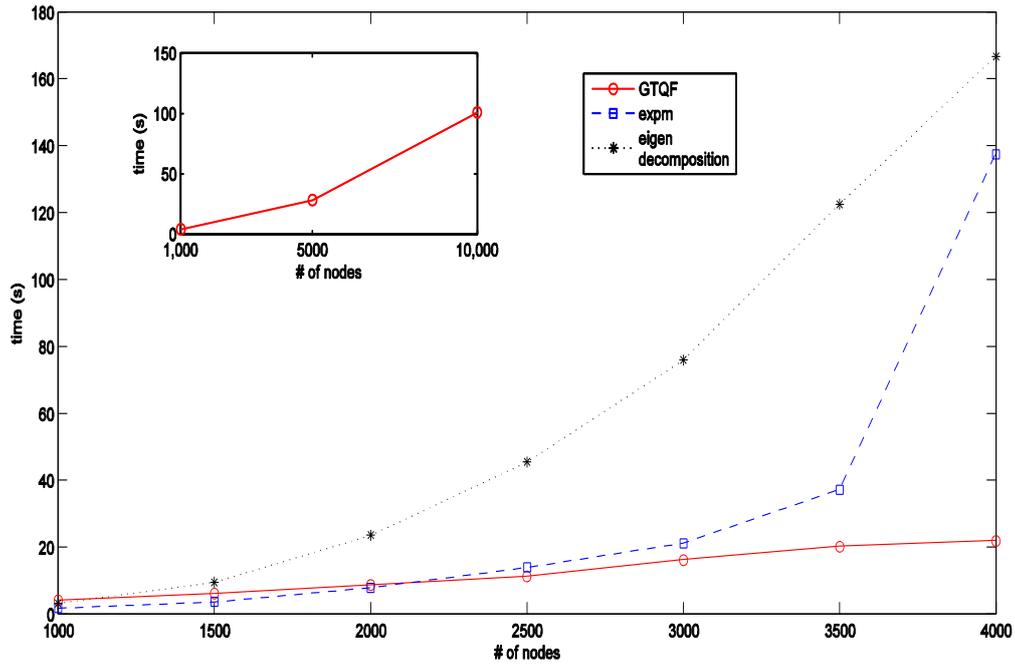

FIG. 14 (color online). (Main) Plot of the time in seconds ($y$-axis) for computing the Estrada index of small world matrices of increasing size. The number of nodes $N$ is on the $x$-axis. In blue we show the time for the Matlab '*expm*' function, in black the time using the eigendecomposition, and in red the time for estimating the trace using five iterations of the Lanczos algorithm for each node. (Insert) Plot of the time in seconds for computing the Estrada index of small world matrices of size $n = 1000, 5000, 10000$.

## VIII. CONCLUSIONS AND PERSPECTIVES

In recent years, the notion of communicability has become increasingly important in the analysis of complex networks. It plays a prominent role in understanding network properties at the micro-, meso-, and macroscopic level, as well as across multiple scales. Measures of the communicability between nodes, or sets of nodes, can be used to construct community detection algorithms, to quantify graph bipartivity, to analyze the spread or rumors, to identify bottlenecks, to reveal the dynamics between agents in a social conflict, and so forth. Moreover, centrality measures based on self-communicability (i.e., subgraph centrality) have proved useful in analyzing the structure of complex networks arising in a variety of fields. A few of these applications have been reviewed here, as well as generalizations and improvements by a number of researchers.

Physical models based on oscillator networks can help justifying and understanding the communicability functions of networks. A variety of models have been analyzed and discussed in this paper, including classical and quantum-mechanical ones. Calculations with these models can provide useful insights on the notion of communicability in various situations; for instance, the use of negative absolute temperatures readily admits an elegant interpretation in the context of complex network analysis.

All the communicability measures reviewed in this paper are expressed in terms of walks between nodes in the graph representing the network. Counting these walks and assigning weights to them so as to penalize longer walks naturally leads to matrix power series and hence to analytic functions of graph matrices such as the adjacency or the Laplacian matrix. A wealth of mathematical and algorithmic knowledge on matrix functions can be utilized to compute communicability functions efficiently in the case of large and sparse networks. Moreover, existing bounds for the entries of matrix functions can be directly applied to investigate the locality (or its absence) in the network communicability. Interest in communicability based on matrix functions has grown very rapidly since the initial proposal (Estrada and Hatano, 2008) and is now an active area of research worldwide. There are still several open questions that have been mentioned or analyzed in this review. Others will appear from the systematic use of the communicability functions in the different application fields. All of them will guarantee the continuity of research in this topic for the next few years.

In closing, the communicability is likely to become an essential ingredient in both the theoretical and the practical analysis of networks, and exciting developments will surely take place in the years ahead.

**Acknowledgement**

NH and EE are partially supported by Aihara Innovative Mathematical Modeling Project, the Japan Society for the Promotion of Science (JSPS) through the "Funding Program for World-Leading Innovative R&D on Science and Technology (FIRST Program)," initiated by the Council for Science and Technology Policy (CSTP). EE was also partially support by EPRSC grant "MOLTEN: Mathematics Of Large Technological Evolving Networks". MB was supported by National Science Foundation grants DMS-0810862 and DMS-1115692.